\documentclass[nohyper,12pt,letterpaper]{JHEP3}
\usepackage[dvips]{epsfig}
\usepackage{amsfonts,amssymb}

\newcommand{\be}{\begin{equation}}
\newcommand{\ee}{\end{equation}}
\newcommand{\ben}{\begin{displaymath}}
\newcommand{\een}{\end{displaymath}}
\newcommand{\bea}{\begin{eqnarray}}
\newcommand{\eea}{\end{eqnarray}}
\newcommand{\bean}{\begin{eqnarray*}}
\newcommand{\eean}{\end{eqnarray*}}

\def\ap {\alpha'}

\def\s {\sigma}

\newcommand{\ads}[1]{\mbox{${AdS}_{#1}$}}
\newcommand{\adss}[2]{\mbox{$AdS_{#1}\times {S}^{#2}$}}

\newcommand{\bra}[1]{\mbox{$\langle #1 |$}}
\newcommand{\ket}[1]{\mbox{$| #1 \rangle$}}

\newcommand{\eg}{{\it e.g.}}
\newcommand{\ie}{{\it i.e.}}

\newcommand{\commentout}[1]{}

\newcommand{\beq}{\begin{equation}}
\newcommand{\eeq}{\end{equation}}
\newcommand{\beqr}{\begin{displaymath}}
\newcommand{\eeqr}{\end{displaymath}}
\newcommand{\beqa}{\begin{eqnarray}}
\newcommand{\eeqa}{\end{eqnarray}}
\newcommand{\beqar}{\begin{eqnarray*}}
\newcommand{\eeqar}{\end{eqnarray*}}
\newcommand{\m}{\mu}

\newcommand{\cN}{{\cal N}}
\newcommand{\cD}{{\cal D}}

\newcommand{\cA}{{\cal A}}

\newcommand{\half}{\ensuremath{\frac{1}{2}}}

\newcommand{\N}[1]{\ensuremath{\cN=#1}}

\newcommand{\ei}{\ensuremath{\varepsilon_i}}
\newcommand{\sz}{\ensuremath{\sigma_0}}
\newcommand{\NC}{\ensuremath{\mathrm{N}}}
\newcommand{\NN}{\ensuremath{\NC^{rs}_{mn}}}
\newcommand{\NNa}{\ensuremath{\NC^{rs}_{mn}(\ei,\sz)}}
\newcommand{\ops}{\ensuremath{1+\sst}}
\newcommand{\sst}{\ensuremath{\sin\frac{\sz}{2}}}
\newcommand{\cst}{\ensuremath{\cos\frac{\sz}{2}}}
\newcommand{\Plm}{\ensuremath{\mathrm{P}^l_m(\cos\sz)}}
\newcommand{\sign}{\ensuremath{\mathrm{sign}}}
\newcommand{\Pp}{\ensuremath{\hat{P}}}
\newcommand{\Id}{\ensuremath{\mathbb{I}}}

\title{\LARGE Planar diagrams in light-cone gauge}

\author{Martin Kruczenski\footnote{After August 15th, at: Department of Physics, Purdue University, 525 Northwestern Avenue, W. Lafayette, IN 47907-2036.} \\
        Department of Physics, Princeton University \\
        Princeton, NJ 08544.

E-mail: \email{martink@princeton.edu}}

\abstract{We consider the open string vacuum amplitude determining the interaction between a stack of N D3-branes and a single probe brane.
When using light cone gauge, it is clear that the sum of planar diagrams (relevant in the large-N limit) is described by the free propagation 
of a closed string. A naive calculation suggests that the Hamiltonian of the closed string is of the form $H=H_0-g_s N \Pp$. 
 The same form of the Hamiltonian follows from considering the bosonic part of the closed string action propagating in the full D3-brane
background suggesting the naive calculation captures the correct information. Further, we compute explicitly $\Pp$ from the open string 
side in the bosonic sector and show that, in a certain limit, the result agrees with the closed string expectations up to extra terms due 
to the fact that we ignored the fermionic sector. 

 We briefly discuss extensions of the results to the superstring and to the sum of planar diagrams in field theory. In particular we argue that
the calculations seem valid whenever one can define a $(\sigma\leftrightarrow\tau)$  dual Hamiltonian in the world-sheet which in principle
does not require the existence of a string action. This seems more generic than the existence of a string dual in the large-N limit.
}


\begin{document}

\section{Introduction}
\label{intro}

 One of the most promising approaches for understanding QCD in the infrared, strong coupling regime,
is the large-N approach proposed by 't Hooft \cite{largeN}. He argued that, particularly when considered
in light cone frame, a theory with $SU(N)$ symmetry and fields in the adjoint looks similar to a string 
theory (also in light cone frame). Although this was a beautiful idea, the development of string theory 
was largely unrelated to it until Polchinski \cite{Polchinski:1995mt} found certain new objects in string theory, 
namely D-branes, that had a description in terms of open strings ending on them or, alternatively, in terms of closed 
strings propagating in certain supergravity backgrounds. Since the low energy limit of the open string theories living on $N$ D-branes is a gauge theory 
with gauge group $SU(N)$ and the closed string description was valid when $N$ was large, the relation was reminiscent of
't Hooft's large-$N$ approach. This connection was understood by Maldacena \cite{malda} who used it to find the first 
concrete example of the relation between the large-N limit of a gauge theory and a string theory. This relation is 
known as the AdS/CFT correspondence and appears as a fundamental step towards understanding
't Hooft's original proposal. In the standard example, the low energy description of a stack of D3 branes leads to a 
relation between $\N{4}$ SYM and $IIB$ string theory on \adss{5}{5}. However, this result is reached indirectly, so that 
it is not clear how to implement the initial idea that one could derive the string Hamiltonian from the field theory. 
In this paper we try to shed some light on this problem.

 To understand what happens, it seems easier to embed the $SU(N)$, $\N{4}$ theory in an open string theory, $IIB$ theory 
on the presence of $N$ $D3$-branes, namely, to go back one step before AdS/CFT is derived. As argued by Polchinski \cite{Polchinski:1995mt}, 
at lowest order, the interaction between a stack of D3 branes and a probe brane is given by the vacuum amplitude of an open string with an end 
on the stack of D3-branes and another in the probe brane.  This vacuum amplitude has an alternative interpretation as a closed string
emerging from the probe brane in a state usually called a boundary state and then being absorbed by the stack of branes also in a 
boundary state (see fig.\ref{DDint}). If the number $N$ of D3 branes is very large then one should replace the stack of D3 branes by a supergravity background
in which the closed string propagates. The interaction is now given by the action of the probe brane in such background.

\FIGURE{\epsfig{file=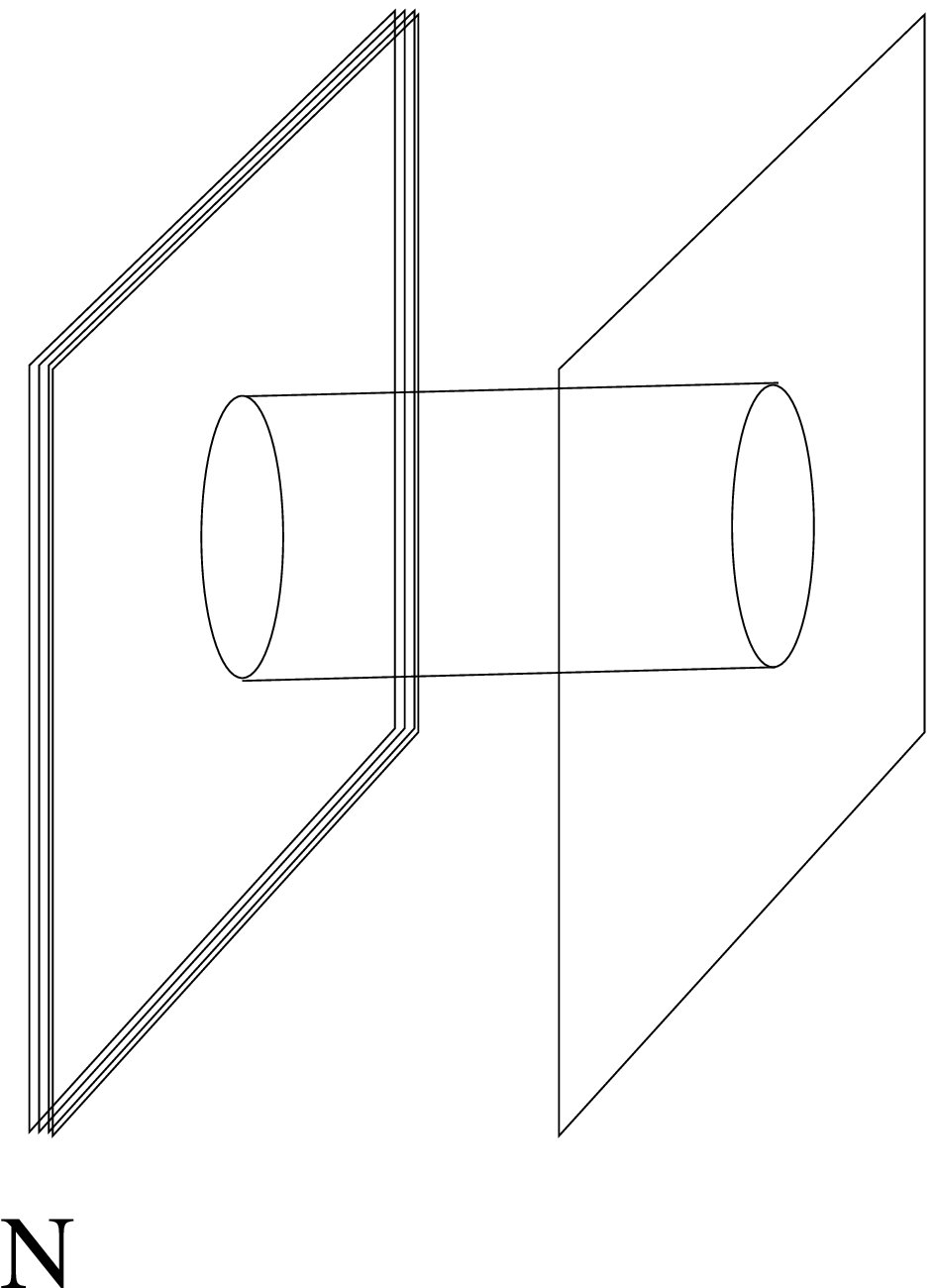, height=6cm}
\caption{The interaction between a stack of $N$ D-branes and a probe brane is given, at lowest order, by a one-loop open string diagram,
or equivalently by a single closed string interchange.}
\label{DDint}
}
 
 In the original picture the supergravity background should appear when summing all planar corrections to the vacuum amplitude.
From the closed string point of view we shall see that, at least in a naive treatment of the problem, the sum over planar diagrams 
is given by the same calculation as the one-loop calculation. Namely, the closed string emerges from the probe brane in a boundary state and 
is absorbed by the stack of branes also in a boundary state. The only difference is that the propagation of the closed string is determined by a  
new Hamiltonian different from simple propagation in flat space. It contains an extra piece that can be described as the operator 
that inserts a hole in the world-sheet. Equivalently, this operator can also be seen as describing the scattering of closed strings
from D-branes. A natural question is therefore if this modified Hamiltonian describes the propagation of a closed string in a modified background. 

 To answer this question we explicitly compute the Hamiltonian and show that in a certain limit it agrees with the expectations.
The comparison is done by studying only the bosonic sector of the theory. So we get extra terms terms due to the presence of the tachyon.
We leave the study of the supersymmetric case for future work. A problem is that it cannot be compared with the string action in the 
full D3 background since that is not known. What one knows is the propagation in the near-horizon (\adss{5}{5}) limit~\cite{Metsaev:1998it}. 
For that we should compare with the field theory limit of the string Hamiltonian. It seems easier in that case to derive the string action 
directly from the \N{4} SYM theory in light cone gauge, since, in that gauge, the theory is greatly simplified \cite{Brink:1982pd}. We expect to report
on this in future work.  

 We should also note that the treatment of the planar diagrams we give here appears as a naive first step since we ignore the presence of
potential divergences that one might have to subtract giving rise to extra terms in the closed string Hamiltonian. In spite of that we
obtained partial agreement with the supergravity background suggesting that at least in certain limits such corrections might be absent.
 We also note that even if those correction are present, the calculation of the operator $\Pp$ that we perform here is interesting in itself
since it seems to contain all the information about the background. 

 Some recent work dealing with deriving a world-sheet description of a gauge theory are \cite{Bardakci:2001cn,Clark:2003wk}. In the first,
a representation in terms of a spin system followed by a mean-field approach is proposed to obtain a world-sheet action and in the second 
a representation of a free field theory in terms of strings is discussed. In the context of the AdS/CFT correspondence a relation between
the Schwinger parametrization of Feynman diagrams and particles propagating in \ads{5} space was discussed in \cite{Gopakumar:2003ns}. A more
detailed analysis of this proposal including various checks can be found in~\cite{Aharony:2006th}.

 Gauge theories in the light cone have been studied in detail \cite{Brodsky:1997de}.  More recent work in that respect 
is \cite{Thorn:2005ak,Belitsky:2004sc}\footnote{I am grateful to L. Brink for pointing out the recent work 
of C. Thorn}. In \cite{Thorn:2005ak}, loop calculations are discussed and in \cite{Belitsky:2004sc} the formulation 
of \N{4} in light cone gauge \cite{Brink:1982pd} is used to compute conformal dimensions of various operators.
 
 String theory in light cone gauge is also very well studied \cite{GSW}. In the case of the superstring that will interest 
us in part of the paper, light cone gauge was an important method used to construct the theory \cite{Green:1983hw, Green:1984fu}.
 More recently, the study of the interaction of closed strings in light cone gauge has also played a role in AdS/CFT \cite{Spradlin:2002ar}.

 The idea of defining a ``hole'' operator was already considered for example in \cite{Polyakov:2001af}. As explained there, it is not clear that
it gives rise to a local world-sheet Hamiltonian and therefore to a dual string theory. 

 All these works, including what we present here, suggest that using light-cone frame is indeed an appropriate framework for understanding the sum 
of planar diagrams, as was envisioned by 't Hooft in~\cite{largeN}. A difference with previous work however is that, for our purpose, we do not need 
a string description of the large-N limit, but only a Hamiltonian description in the dual ($\sigma\leftrightarrow\tau$) channel. In particular the
Hamiltonian we get is non-local and therefore it does not have a clear interpretation as a string Hamiltonian. Nevertheless it can still be useful
to understand the properties of planar diagrams. 

 To finish, let us clarify that, in a generic field theory, light cone frame refers to using light-like coordinates $X^{\pm}$ to quantize
the theory whereas light cone gauge refers to taking $A_+=0$ in a gauge theory. In string theory light cone gauge refers to taking $X^+=\tau$ and in the
field theory limit is related to both light cone frame/gauge.

\section{Planar diagrams in light cone gauge}
\label{planar}

 As depicted in fig.\ref{DDint}, the interaction between two D-branes can be computed, at lowest order as a one-loop vacuum diagram 
of an open string stretching between them or as the propagation of a closed string emerging
from one D-brane and disappearing in the other. The initial and final states of the closed
string are the so-called boundary states corresponding to the D-branes in question. The two 
calculations are related by an interchange between $\sigma$ and $\tau$, the world-sheet coordinates.
 In the path integral approach both calculations are the same, but if we use a Hamiltonian approach 
they differ in which world-sheet direction we take to be time and which to be space. 

 Suppose now that we take time to be such that we have open string states and do the computation in  
light-cone gauge. In order to do that, we choose a direction $X$ parallel to the brane and define $X^{\pm}=(x\pm t)$.
We use conformal gauge by fixing the world-sheet metric to be equal to the identity and use a residual symmetry to 
take $X^+=\tau$ identifying world-sheet time with $X^+$. Since we want to compute a partition function, we take 
$\tau$ and therefore $X^+$ to be periodic with period $2\pi$. 

 If we now want to include higher order corrections we should use the three-open string vertex which 
allows for open strings to split and rejoin. A typical diagram looks like the one in figure \ref{fig1}. 
In this section we study the sum of all those diagrams that can be drawn by adding slits to the cylinder as in the 
figure. There are many other diagrams where the open strings cross over each other before reconnecting. 
Those are non-planar and we ignore them here since we are interested in the large-$N$ limit. 

\FIGURE{\epsfig{file=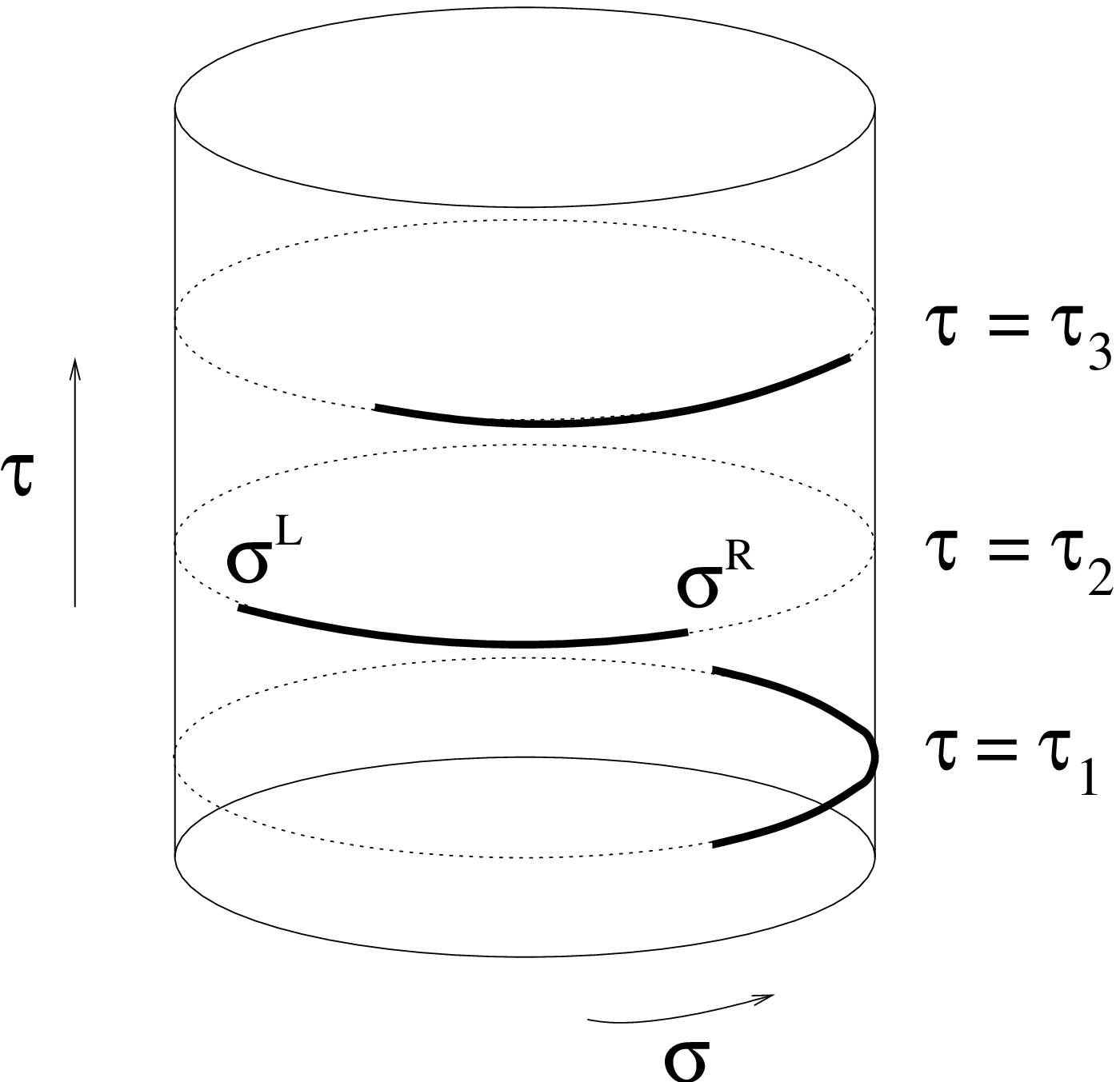, height=10cm}
\caption{Typical diagram appearing in the computation of the open string vacuum amplitude in light cone gauge. In this channel
a single closed string propagates suffering self interactions at certain times.}
\label{fig1}
}
  
 In the path integral approach~\cite{GSW} we should compute, for a given diagram with $n$ holes 
\beq
Z_n = \int \prod_{i=1}^n {d\sigma^L_{i}\,d\sigma^R_{i}\, d\tau_i} \int \mathcal{D}X\, e^{\int d\sigma d\tau \partial X_\mu \bar{\partial} X^{\mu}} ,
\eeq
where the path integral is over configurations that obey appropriate boundary conditions on the slits. We consider the case when 
all those slits are on the D3-branes sitting at the origin. The other D-brane is taken as a probe. We
should integrate over all positions of the slits $(\sigma^L_i,\sigma^R_i,\tau_i)$, where there are two $\sigma$'s and one
$\tau$ for each slit, namely three parameters. The cuts are indistinguishable so the integral should be such that
we do not count configurations related by interchanging cuts. Finally we should weigh the diagram with a 
factor $(g_s N)^n$ ($n$ is the number of slits). 

We make here two important assumptions about this expression. First, that the measure
of integration for the parameters $\sigma^L_i$, $\sigma^R_i$ and $\tau_i$ is completely determined by the path integral without any 
extra functions. Second, that there is no $n$ dependent factor in front (other than a power $A^n$ that can be absorbed in the coupling constant).
 These assumptions can be relaxed slightly as we discussed later but without them we cannot do the rest of the calculation. It is not clear
to us if the relative weight of the string theory diagrams has been studied carefully enough to know if these assumptions give rise to consistent
amplitudes. On the other hand, the comparison we make later in the paper between the open string diagrams and propagation of a closed string
suggest that they are reasonable. 

 If we now use a Hamiltonian approach but in the closed string channel we see that we always have only one closed string\footnote{This
closed string is wrapped around the direction $X^+$ that we took to be periodic.},
namely it does not split, so it is a ``free'' closed string propagator. The only caveat is that at certain times
$\tau_i$ the closed string suffers a self-interaction. We can define an operator $\Pp_{\sigma^L_i\sigma^R_i}$ that propagates the closed string
from $\tau_i-\epsilon$ to $\tau_i+\epsilon$, ($\epsilon\rightarrow 0$). Using this operator we can rewrite $Z_n$ as
\beq
Z_n = \int_{t_i<\tau_1<\ldots <\tau_n<t_f} \prod_{i=1}^n {d\sigma^L_id\sigma^R_id\tau_i} \, \bra{f}\, e^{-iH_0(t_f-t_{n})}\, \Pp_{\sigma^L_n\sigma^R_n}\,
e^{-H_0(t_n-t_{n-1})}\ldots \Pp_{\sigma^L_1\sigma^R_1}\, e^{-H_0(t_1-t_{i})}\, \ket{i} ,
\eeq
The initial and final states $\ket{i}$, $\ket{f}$ are boundary states and $H_0$ is the free closed string Hamiltonian.  
Since the only dependence on $\sigma_{L,R}$ is in the $\Pp$'s we can define a new operator 
\beq
\Pp = \int d{\sigma^L} d {\sigma^R} \Pp_{\sigma^L\sigma^R} .
\eeq
 The only dependence on $\tau_i$ is in the free propagators. We can define new variables $\xi_i=t_i-t_{i-1}\ge 0, i=1\ldots n+1$ and rewrite
everything as
\beq
 Z = \int_0^\infty \prod_{i=1}^{n+1} d\xi_i\, \delta\!\left(\sum\xi_i - (t_f-t_i)\right) \bra{f} e^{-H_0\xi_{n+1}} \Pp
e^{-H_0\xi_{n}}\ldots \Pp e^{-H_0\xi_2} \Pp e^{-H_0\xi_1} \ket{i} .
\eeq
 The total vacuum amplitude, namely the sum over all planar diagrams is given by
\beqa
Z_T &=& \sum_n (g_sN)^n \int_0^\infty \prod_{i=1}^{n+1} d\xi_i\, \delta\!\left(\sum\xi_i - (t_f-t_i)\right) \bra{f} e^{-H_0\xi_{n+1}} 
\Pp e^{-H_0\xi_{n}}  \ldots \Pp e^{-H_0\xi_1} \ket{i} \nonumber \\
    &=& \bra{f}  \sum_n (g_sN)^n \int_{-\infty}^{\infty}  \frac{d\omega}{2\pi} \int_0^\infty \prod_{i=1}^{n+1} e^{i\omega(\sum\xi_i-(t_f-t_i))} e^{-H_0\xi_{n+1}} \Pp 
e^{-H_0\xi_{n}}\ldots \Pp e^{-H_0\xi_2} \Pp e^{-H_0\xi_1} \ket{i} \nonumber\\
    &=& \bra{f} \int_{-\infty}^{\infty} 
          \frac{d\omega}{2\pi} e^{i\omega(t_f-t_i)} \sum_n (g_sN)^n \frac{1}{H_0-i\omega} \left[\Pp \frac{1}{H_0-i\omega} \right]^n \ket{i} \\
    &=& \bra{f}  \int_{-\infty}^{\infty} \frac{d\omega}{2\pi} e^{i\omega(t_f-t_i)} \frac{1}{H_0 - (g_sN) \Pp -i\omega} \ket{i} \nonumber\\
    &=& \bra{f} \int_0^{\infty} d\xi \int_{-\infty}^{\infty}  \frac{d\omega}{2\pi} e^{i\omega(t_f-t_i)+i\omega\xi-(H_0-(g_sN)\Pp)\xi} \ket{i} \nonumber\\
    &=& \bra{f} \int_0^{\infty} d\xi \delta(\xi- (t_f-t_i)) e^{-(H_0-(g_sN)\Pp)\xi} \ket{i} \nonumber\\
    &=& \bra{f} e^{-(H_0-(g_sN)\Pp)(t_f-t_i)} \ket{i} \nonumber.
\eeqa
 These elementary manipulations lead to the result that the partition function can indeed be computed as a free string propagation (in the
sense that the string does not split) but with a modified Hamiltonian $H=H_0-\lambda \Pp$ where $\lambda=g_sN$.
 It is clear that this is independent of the initial and final states so we can ignore those and simply study closed strings
with Hamiltonian $H$. Such Hamiltonian determines the ``dual'' closed string picture to the open string one. It is a bit peculiar
because $\Pp$ is a non-local operator in the world-sheet so the interpretation of $H$ as a string Hamiltonian is unclear for now. What is 
certain is that it describes a Hamiltonian  of a non-local one-dimensional system that contains the information about the sum of all planar
diagrams in the open string channel. As part of that it also contains information about the sum of all planar diagrams in the low energy 
gauge theory. 

 We can now go back to the assumptions we made. We see that extra $\sigma_i^{L,R}$-dependent functions are acceptable as long as they factorize such that
they can be absorbed in the operator $\Pp$. However we need to have locality in $\tau$, namely, if we cut the diagram at a certain value of $\tau$ and
introduce an identity, the diagram should factorize in two independent pieces. This means that if we take $\tau_1<\tau_2<\tau_3$, propagation from $\tau_1$
to $\tau_3$ is the same as propagation from $\tau_1$ to $\tau_2$ and then from $\tau_2$ to $\tau_3$. Another potential problem is the presence of an $n$-dependent    
coefficient in front of the diagram. In that case, instead of an exponential, the sum will give another function of $H_0$ and $\lambda\Pp$. This could still
be tractable but certainly more cumbersome. 

 Even if these assumptions are correct, the expression for the amplitude might still not be well defined. In the superstring one has to insert extra operators
at the end points of the slit where the world-sheet is singular. These operators contract among themselves when the slits come close to each other, giving rise
to singularities that need to be subtracted. What this means is that in the string Hamiltonian there are vertices of order higher than 
cubic~\cite{Greensite:1987hm,Green:1983hw,Green:1984fu} making the 
light cone superstring possibly ill defined\footnote{I am grateful to N. Berkovits for explaining this to me.}. Furthermore, when two slits come together 
there can be singularities already in the bosonic string\footnote{I am grateful to J. Maldacena for emphasizing this point.}. When the slits are small, these 
singularities are of the type that appear when two world-sheet operators come close and are usually avoided by analytic continuation in the momenta of the operators
but it is not clear if that idea can also be used here. When the slits that come together are long, the singularity comes from an open string tachyon propagating 
between the slits which seems to be an unavoidable problem in the bosonic string\footnote{I am grateful to I. Klebanov for pointing this out.}. In the superstring 
it should go away.   

In any case, all these considerations show that the calculation we did is too naive and one might expect to get corrections to $\Pp$ of higher order in $\lambda$. 

 In fact, at first, this sounds as a very reasonable possibility since the closed string Hamiltonian should describe propagation of the closed string 
in the full D-brane background which has a non-trivial dependence on $\lambda$. However, in the next section we show that, in spite of the complicated
background, in this gauge, the propagation of the closed string in the full D-brane background is described by a Hamiltonian first order in $\lambda$.
 This adds plausibility to the idea that the sum exponentiates and that the naive calculation captures some important physics. Moreover, in later 
sections we compute $\Pp$ and compare with the expectation from closed string propagation and find agreement up to terms that we attribute to the fact 
that we do so only for the bosonic sector of the theory. 

 An optimistic interpretation would be that we can view the exponentiation as a particular way of regularizing the divergences we discussed and therefore
is a consistent result for the sum of planar diagrams. 

 Before going into the calculation it is interesting to discuss what happens if we want to compute a scattering amplitude instead of a vacuum amplitude.
 This is clarified in fig.\ref{fig2} where we see that, after interchanging $\sigma\leftrightarrow\tau$ we should compute the propagation of an
infinitely long string. This string has the same self-interaction from the slits to which we should add the special cuts that extend to infinite.
 It is clear that the contribution of the slits is the same as before and therefore give rise to the same Hamiltonian. In terms of the 
AdS/CFT correspondence it seems that the infinitely long string should be interpreted, in some sense as a string ending in the boundary. 
 In this paper we do not discuss further the calculation of scattering amplitudes and concentrate in the diagrams of fig.\ref{fig1}. 
Note however, that whether we compute scattering amplitudes or vacuum amplitudes, the closed strings we need to consider are ``long'' since
$X^+=\sigma$. We do not need to study point-like strings, namely those such that all coordinates are independent of $\sigma$.

\FIGURE{\epsfig{file=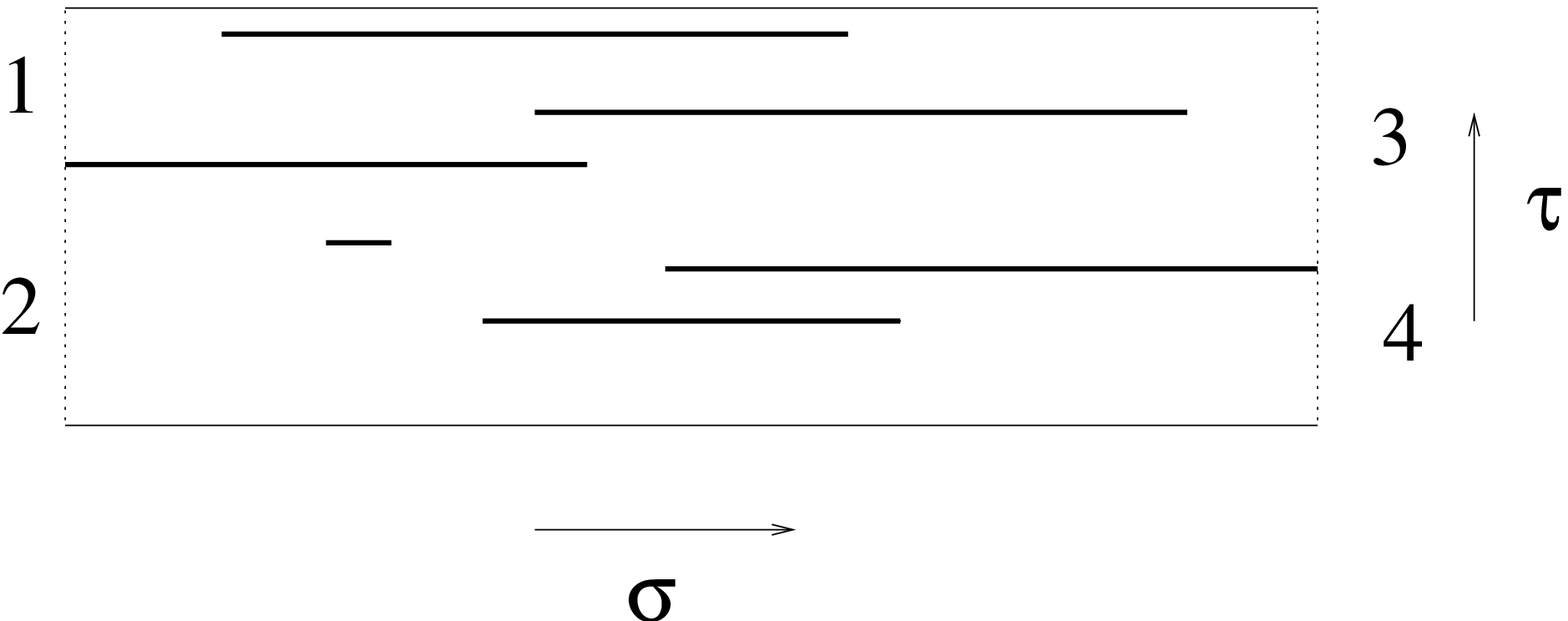, width=10cm}
\caption{Typical diagram appearing in the computation of a two particle scattering amplitude ($1+2\rightarrow 3+4$) in light cone gauge. 
We flipped $\sigma$ and $\tau$ with respect to the usual open string channel parameterization. In this channel an infinitely long string propagates 
having self interactions at certain times.}
\label{fig2}
}

\section{Closed strings in the D3-brane background }
\label{bkg}

 In this section we consider the bosonic part of the action of a string moving in the full D3-brane background. The metric is
\beq 
ds^2 = \frac{1}{\sqrt{f}} \left(dX^+dX^-+ dX^2\right) + \sqrt{f} dY^2, \ \ \ \ \ f = 1+ 4\pi\ap{}^2\frac{g_sN}{Y^4} ,
\eeq
where $X^{\pm},X$ denote the coordinates parallel to the brane and $Y$ those perpendicular. The Polyakov action in such 
background is 
\beq
S = \frac{1}{4\pi\ap} \int d\sigma d\tau \sqrt{h} h^{ab} G_{\mu\nu} \partial_aX^{\mu} \partial_b X^{\nu} .
\eeq
 We expect that the match to the string side happens in light cone gauge since we do the open string calculation there.
 However, in a curved background things are slightly more complicated and we are going to follow the ideas in \cite{Metsaev:2000yu,Polchinski:2001ju} 
where the similar case of \ads{5} was considered. 

 As indicated in \cite{Polchinski:2001ju}, we can fix the gauge by taking $h_{01}=0$, $X^+=\tau$. 
The action reduces to
\beq
 S = \frac{1}{4\pi\ap} \int d\sigma d\tau  \left[ E(\dot{X}^-+\dot{X}^2) - \frac{1}{f} EX'^2+Ef\dot{Y}^2 - \frac{1}{E}Y'^2 \right] ,
\eeq
 where $E = \sqrt{-\frac{h_{11}}{h_{00}f}}$. The equation of motion for $X^-$ implies that $E$ is a function of $\sigma$ only
so we can set it to $1$ by redefining $\sigma$ accordingly. We end up with an action
\beq
 S = \frac{1}{4\pi\ap} \int d\sigma d\tau  \left[ \dot{X}^2 - \frac{1}{f} X'^2+ f\dot{Y}^2 - Y'^2 \right] .
\eeq
 Now we want to compute the Hamiltonian. It turns out to be
\beq
 H = \frac{1}{4\pi\ap} \int d\sigma \left[ (2\pi\ap\Pi_X)^2 + \frac{1}{f} (2\pi\ap\Pi_Y)^2 + \frac{1}{f} X'^2 + Y'^2 \right] ,
\eeq
which is clearly not of the form $H=H_0-\lambda \Pp$ so we seem to have failed. However we should now remember that in the open string we fixed
$X^+=\tau$ and, when going to the closed string channel, that becomes $\sigma$, the spatial direction. If we interchange $\sigma$
and $\tau$ in the action we get (we also introduce an overall sign while doing that):
\beq
 S = \frac{1}{4\pi\ap} \int d\sigma d\tau  \left[ -X'^2 + \frac{1}{f} \dot{X}^2- fY'^2 + \dot{Y}^2 \right] .
\eeq
Although nothing dramatic happens from the point of view of the action, if we compute the Hamiltonian now we get
\beq
 H = \frac{1}{4\pi\ap} \int d\sigma \left\{ (2\pi\ap\Pi_Y)^2 + X'^2 + f \left[(2\pi\ap\Pi_X)^2+Y'^2\right] \right\} .
\eeq
Recalling that $f=1+4\pi\ap^2\frac{g_sN}{Y^4}$ we get
\beq
 H = \frac{1}{4\pi\ap} \int d\sigma \left\{ (2\pi\ap\Pi_Y)^2 + X'^2 + (2\pi\ap\Pi_X)^2 + Y'^2 + \frac{4\pi\ap^2g_sN}{Y^4} \left[(2\pi\ap\Pi_X)^2+Y'^2\right] \right\} ,
\label{HfromD}
\eeq
 which indeed is of the form $H-\lambda \Pp$ with 
\beq
\Pp=-\ap \int d\sigma  \frac{1}{Y^4} \left[(2\pi\ap\Pi_X)^2+Y'^2\right] .
\label{Pbulk}
\eeq
 So in this gauge, closely related to light-cone gauge, that we can call $\sigma$-gauge (because $X^+=\sigma$) the Hamiltonian has
the desired form. To check units we should use that $\sigma$, $Y$ have units of length, $\ap$ of length squared and $\Pi_X$ of $[\mbox{length}]^{-1}$. 
One should note that this result is true for branes of any dimensionality even if here we are interested only in D3-branes.
 In the D3-brane case, the Maldacena (near-horizon) limit, which can loosely be described as ``dropping the $1$'' in the function $f$, leads 
to the Hamiltonian 
\beq
 H = \frac{1}{4\pi\ap} \int d\sigma \left\{ (2\pi\ap\Pi_Y)^2 + X'^2 + \frac{4\pi\ap^2g_sN}{Y^4} \left[(2\pi\ap\Pi_X)^2+Y'^2\right] \right\} ,
\label{ftlimbulk}
\eeq
which describes propagation of closed strings in \ads{5}(see \eg\ \cite{Polchinski:2001ju} for a related Hamiltonian). Note that this Hamiltonian 
scales as $\mu^2$ under the scale transformation $X\rightarrow \mu X$, $Y\rightarrow Y/\mu$ whereas the full Hamiltonian (\ref{HfromD}) does not.    

 Going back to (\ref{HfromD}), it remains to be seen if the operator $\Pp$ that we found here is the same as the one we derived from the open string side. 
In the next section we compute $\Pp$ from the open string point of view and make the comparison.

\section{The hole operator $\Pp$}
\label{Pcalc}

In this section we compute the operator $\Pp$. This amounts to finding the relation between the state of the closed string just before 
and just after a time where we insert a hole or slit. In the case of all Dirichlet boundary conditions, this problem was solved by 
Green and Wai \cite{Green:1994ix}. Since we need also to consider Neumann boundary conditions and our expressions are slightly simpler 
we include a detailed account of the calculation \footnote{In fact we learned about \cite{Green:1994ix} after we had finished the calculation
which partially explains why we obtained different looking expressions}.

\FIGURE{\epsfig{file=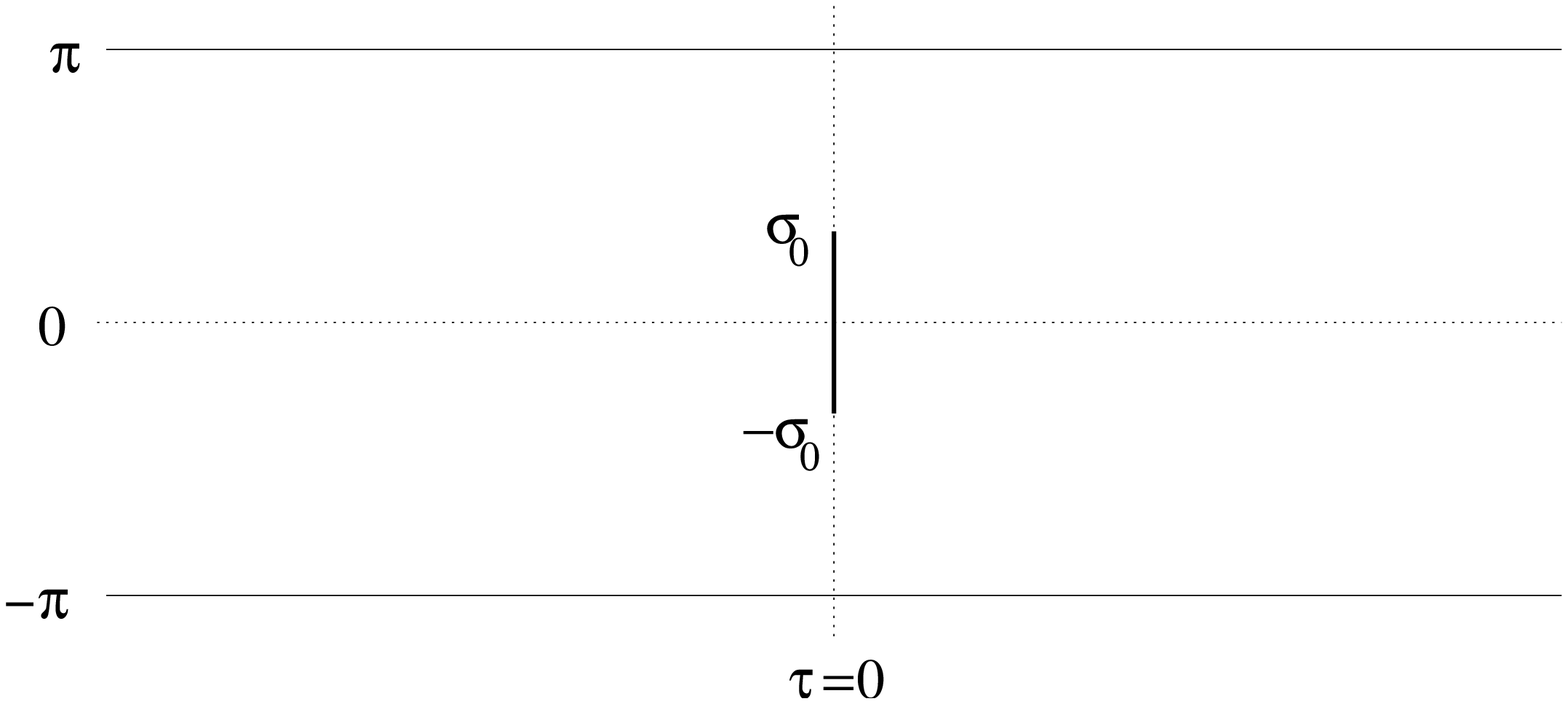, height=6cm}
\caption{Diagram defining the operator $\Pp$. The upper and lower lines are identified so this diagram describes the propagation of a closed string. 
It propagates freely from  $\tau=-\infty$ to $\tau=0$ where we apply $\Pp$. Then it propagates freely again to $\tau=+\infty$. Removing the free pieces 
from the amplitude we get the operator $\Pp$. }
\label{slitfig}
}

\subsection{Computation of $\Pp$.}
 
 To understand the definition of $\Pp$ it is useful to include it in an infinite cylinder as in fig.\ref{slitfig}. This diagram read from left to right,
can be thought as a closed string propagating freely up to $\tau=0$ where we apply $\Pp$. After that we let it evolve freely again. In fact this is  
actually the calculation we should do to compute closed string scattering from a D-brane for generic closed string states. Although this interpretation 
clarifies the meaning of $\Pp$, obtaining a precise expression for it is a rather lengthy calculation (see \eg\ \cite{GSW}) that we divide in some steps. 
Summarizing: first we use a representation of $\Pp$ as a two string vertex state that has to satisfy certain properties from continuity of the coordinates 
and boundary conditions on the slit. After that we show that these conditions have a solution in the oscillator basis in terms of some undetermined 
Neumann coefficients. Finally, using the relation to the scattering amplitude we find these coefficients. 

\subsubsection{$\Pp$ as a two closed string vertex}

 We are going to describe the operators in terms of a vertex state defined in the product of two single 
strings Hilbert spaces. More precisely we take
\beq
\ket{V} = \sum_{\ket{1}\ket{2}} \bra{2}\Pp\ket{1}\ \ket{1}\otimes \ket{2} ,
\eeq
 where $\ket{1}\otimes \ket{2}$ is a state in the product Hilbert space of two strings and $\bra{2}\Pp\ket{1}$ is the matrix element 
of $\Pp$ when considering the two states as being states of a single string. The state $\ket{V}$ so defined has the property
\beq
 \Pp \ket{1} = \bra{1}V\rangle .
\eeq
 This seems a bit convoluted but is standard and helps simplifying some expressions. In fact what we do in this section 
is essentially to follow Chapter 11 in \cite{GSW}. 

 One way to compute $\ket{V}$, is simply to observe that the states before and after the self interaction, are related by continuity and 
conservation of momenta in the region $\sz<\s<2\pi-\sz$ where there is no slit whereas they are projected to the appropriate boundary 
conditions on the slit $-\sz<\s<\sz$.

The state $\ket{V}$  is then required to satisfy
\beqa
 \left(X_1(\sigma,\tau=0) - X_2(\sigma,\tau=0)\right) \ket{V} &=& 0 , \ \ \sz \le |\sigma| \le \pi ,\\
 \left(P_1(\sigma,\tau=0) + P_2(\sigma,\tau=0)\right) \ket{V} &=& 0 , \ \ \sz \le |\sigma| \le \pi ,\\
  X_1(\sigma,\tau=0) \ket{V} &=& 0 , \ \ |\sigma|\le \sz ,\\
  X_2(\sigma,\tau=0) \ket{V} &=& 0 , \ \ |\sigma|\le \sz ,
\eeqa
for Dirichlet bdy. conditions and
\beqa
 \left(X_1(\sigma,\tau=0) - X_2(\sigma,\tau=0)\right) \ket{V} &=& 0 , \ \ \sz \le |\sigma| \le \pi ,\\
 \left(P_1(\sigma,\tau=0) + P_2(\sigma,\tau=0)\right) \ket{V} &=& 0 , \ \ \sz \le |\sigma| \le \pi ,\\
  P_1(\sigma,\tau=0) \ket{V} &=& 0 , \ \ |\sigma|\le \sz ,\\
  P_2(\sigma,\tau=0) \ket{V} &=& 0 , \ \ |\sigma|\le \sz ,
\eeqa
for Neumann (since $P=\partial_\tau X$). 
This can be rewritten in a simpler way as:
\beqa
 \left(X_1(\sigma) - X_2(\sigma)\right) \ket{V} &=& 0 , \ \ -\pi\le \sigma\le \pi ,\\
 \left(X_1(\sigma) + X_2(\sigma)\right) \ket{V} &=& 0 , \ \ |\sigma|\le \sz ,\\
 \left(P_1(\sigma) + P_2(\sigma)\right) \ket{V} &=& 0 , \ \ \sz \le |\sigma| \le \pi ,
\eeqa
for Dirichlet and
\beqa
 \left(P_1(\sigma) + P_2(\sigma)\right) \ket{V} &=& 0 , \ \ -\pi\le \sigma\le \pi ,\\
 \left(P_1(\sigma) - P_2(\sigma)\right) \ket{V} &=& 0 , \ \ |\sigma|\le \sz ,\\
 \left(X_1(\sigma) - X_2(\sigma)\right) \ket{V} &=& 0 , \ \ \sz \le |\sigma| \le \pi ,
\eeqa
for Neumann, where we understand all operators are evaluated  at $\tau=0$. 
These equations are enough to find the two-strings vertex state. Before going into that let us make a quick argument to show that these equations can be solved 
and which form the solution has. 

Suppose we define a generalized 
boundary state $\ket{B(\mathcal{O})}$ through the condition $\mathcal{O}(\sigma)\ket{B(\mathcal{O})}=0, \forall \sigma$. It is well 
known how to find such state if $\mathcal{O}=X$ or $\mathcal{O}=P$\cite{DiVecchia:1999rh}. Having that we can construct our desired state as
\beqa
\ket{V}_D &=& e^{i\frac{\pi}{8}\int_{-\sz}^{\sz}(P_1+P_2)^2+(X_1+X_2)^2} \ket{B(X_1-X_2)}\otimes \ket{B(P_1+P_2)} , \\
\ket{V}_N &=& e^{i\frac{\pi}{8}\int_{-\sz}^{\sz}(P_1-P_2)^2+(X_1-X_2)^2} \ket{B(P_1+P_2)}\otimes \ket{B(X_1-X_2)} . 
\eeqa
 The operators in the exponent are such that they rotate $P$ into $-X$ (and $X$ into $P$). This is easily seen since they 
are harmonic oscillator Hamiltonians\footnote{This should not be confused with the Hamiltonian of the string which contains $X'^2$ and
is completely unrelated. We are just using a trick to write the state explicitly.} independent at each value of $\sigma$. The standard 
time evolution of the harmonic oscillator interchanges $X$ and $P$ after one quarter period. Here that is $\frac{\pi}{2}$, there is an extra 
$\frac{1}{4}$ because the Hamiltonian is $h= \half (\tilde{p}^2+\tilde{x}^2)$, with $\tilde{p}=\frac{1}{\sqrt{2}}(P_1\pm P_2)$ and the same 
for $\tilde{x}$. To find the state in the oscillator number basis we can replace
$X$ and $P$ in terms of $a_n$ and $a^{\dagger}_n$ and normal order the exponential (including what comes from writing the boundary states).
 To do that we expand $X$ and $P$ in normal modes as
\beqa
X_r^i &=& x_r^i + \sum_{n\neq 0} \frac{i}{|n|}\left(a_{irn} - a_{ir,-n}^\dagger\right) e^{in\sigma} ,\\
P_r^i &=& \frac{1}{2\pi} \left[a_{ir0}^\dagger +\half \sum_{n\neq 0}\left(a_{inr} + a_{ir,-n}^\dagger\right)  e^{in\sigma} \right] ,
\eeqa
 where $n$ is summed from $-\infty$ to $+\infty$. 
 We extracted the zero mode and defined for convenience $p_0=a_{i0r}^\dagger$. Besides, 
the index $i$ labels the coordinate and the index $r=1,2$ labels the string we consider.
 Also, the commutation relations read
\beq
[a_{irn},a_{jsm}^\dagger] = |n|\, \delta_{ij}\delta_{rs}\,\delta_{mn} .
\eeq
After normal ordering, the result for $\ket{V}$ is of the form
\beq
 \ket{V} = e^{\sum_{rs,imn} N^{rs}_{i,nm} a_{irn}^\dagger a_{ism}^\dagger }   \prod_{i/\ei=+1} \delta(p^i_1+p^i_2) \ket{0}  ,
\eeq 
 where we still have to determine the coefficients $\NNa$. 
 The indices $r,s=1,2$ label the string and $-\infty<m,n<\infty$ the Fourier modes. 
Also, the state $\ket{0}$ is the vacuum of the oscillators with $n\neq 0$. The state depends
on $p_0=a_{i0r}^\dagger$ and we consider such dependence as the wave-function of the state in momentum representation. For the Neumann coordinates,
the wave function has an explicit delta function for the zero modes. This is indicated by the constants $\ei$ which we define to be $+1$ for
directions $i$ satisfying Neumann boundary conditions and $\ei=-1$ for Dirichlet. This allows for more compact expressions for the coefficients 
$N^{rs}_{i,nm}$ which are Dirichlet or Neumann depending on the direction $i$ and should follow after normal ordering the expression one gets for $\ket{V}$.  

 \subsubsection{Equations for the Neumann coefficients}

Instead of doing the normal ordering calculation it is easier to convert the equations for $\ket{V}$ into equations for $\NNa$. 
To write those conditions we use that
\beqa
 X^i_r \ket{V} &=& e^{\Delta_B} \sum_{sm} 2i\left[\sum_n N^{rs}_{i,mn} e^{in\phi}- \sum_{n\neq 0}\frac{1}{2|n|}\delta^{rs}_{m+n} e^{in\phi}\right] 
                   a^{\dagger}_{ism}  \ket{0} ,\\
 P^i_r \ket{V} &=& e^{\Delta_B} \frac{1}{2\pi} \sum_{sm}\left[\delta^{sr} \delta_{m0} +\sum_{n\neq 0} |n| N^{rs}_{i,nm} e^{in\phi} 
                   + \half \delta^{rs}_{m+n} e^{in\phi}\right] a^\dagger_{ism} \ket{0} ,
\eeqa
 where we commuted $X^i_r$ and $P^i_r$ with the exponential $\exp(\Delta_B)$, $\Delta_B=\sum_{rs,mn} N^{rs}_{i,nm} a_{irn}^\dagger a_{ism}^\dagger$ 
Consider now the case of Dirichlet boundary conditions. We get that
\beq
 \left(X_1(\sigma) - X_2(\sigma)\right) \ket{V} = 0 , \ \ -\pi\le \sigma\le \pi ,
\label{firstNeq}
\eeq
is valid if
\beq
\sum_n \left(N^{1s}_{i,nm}-N^{2s}_{i,nm}\right)e^{in\sigma} = \frac{\delta_{m\neq 0}}{2|m|}(\delta^{1s}-\delta^{2s}) e^{-im\sigma}  , 
    \ \ -\pi\le \sigma\le \pi ,
\eeq
 whereas
\beq
 \left(X_1(\sigma) + X_2(\sigma)\right) \ket{V} = 0 , \ \ |\sigma|\le \sz ,
\eeq
is valid if
\beq
 \sum_n \left(N^{1s}_{i,nm}+N^{2s}_{i,nm}\right)e^{in\sigma} =  \frac{\delta_{m\neq 0}}{2|m|}(\delta^{1s}+\delta^{2s}) e^{-im\sigma}  , \ \ |\sigma|\le \sz .
\eeq
 Finally, for
\beq
 \left(P_1(\sigma) + P_2(\sigma)\right) \ket{V} = 0 , \ \ \sz \le |\sigma| \le \pi ,
\eeq
 we need
\beq
 \delta_{m0} + \sum_n |n| \left(N^{1s}_{i,nm}+N^{2s}_{i,nm}\right)e^{in\sigma} + \half e^{-im\sigma}=0, \ \ \sz \le |\sigma| \le \pi .
\eeq
 Similar relations are obtained for the Neumann case. We get that
\beq
 \left(P_1(\sigma) + P_2(\sigma)\right) \ket{V} = 0 , \ \ -\pi\le \sigma\le \pi ,
\eeq
is valid if
\beq
\sum_n |n| \left(N^{1s}_{i,nm}+N^{2s}_{i,nm}\right)e^{in\sigma} = -\half  e^{-im\sigma}  , 
    \ \ -\pi\le \sigma\le \pi , \ \ (m\neq 0),
\eeq
 whereas
\beq
 \left(P_1(\sigma) - P_2(\sigma)\right) \ket{V} = 0 , \ \ |\sigma|\le \sz ,
\eeq
is valid if
\beq
 \sum_n |n| \left(N^{1s}_{i,nm}-N^{2s}_{i,nm}\right)e^{in\sigma} =  -(\delta^{1s}-\delta^{2s})\left[\delta_{m0} + \half \delta_{m\neq 0}  e^{-im\sigma} \right]  
                    , \ \ |\sigma|\le \sz .
\eeq
 Finally, for
\beq
 \left(X_1(\sigma) - X_2(\sigma)\right) \ket{V} = 0 , \ \ \sz \le |\sigma| \le \pi ,
\eeq
 we need
\beq
 \sum_n \left(N^{1s}_{i,nm}-N^{2s}_{i,nm}\right)e^{in\sigma} = \delta_{m\neq 0} \frac{1}{2|m|}(\delta^{1s}-\delta^{2s}) e^{-im\sigma}=0, \ \ \sz \le |\sigma| \le \pi .
\label{lastNeq}
\eeq
In the following we find the solution to the equations~(\ref{firstNeq})-(\ref{lastNeq}).

\subsubsection{Obtaining $\NNa$}

Instead of solving the equations directly we found useful to extract the Neumann coefficients directly from the amplitude in fig.\ref{slitfig} where
we considered the infinite cylinder $-\infty<\tau<\infty$, $-\pi<\s<\pi$, and inserted a slit at $\tau=0$. If we send a closed string state $\bra{1}$  
from $\tau=\tau_i\rightarrow -\infty$, it will evolve with the free closed string Hamiltonian until it reaches $\tau=0$, there we have to apply the 
operator $\Pp$ and then it will keep evolving with the free Hamiltonian.   
 If we compute then the overlap with another state at $\tau=\tau_f\rightarrow \infty$ and extract the exponentials $\exp(-P_-^{1} \tau_i+P_-^2 \tau_f)$ 
(where $P_-^{1,2}$ are the light cone energies of the initial and final states) then we are left with the matrix element $\bra{1} \Pp\ket{2}$. 

 As we mentioned, this amplitude is actually the scattering amplitude of a closed string by a D-brane evaluated between generic closed string states.
Such amplitude is computed as a path integral over all string configurations satisfying prescribed boundary conditions. One boundary condition
is the Neumann or Dirichlet boundary condition at the slit. The other are boundary conditions at $\tau=\pm \infty$ that determine the initial
and final state of the string. Those we take to be Neumann since we are going to represent the state as wave functions in the momentum representation.
 The path integral is quadratic so the only object we need is the Green function on the cylinder with a slit. Given such Green function, standard 
manipulations (see chapter 11 in \cite{GSW}) give for the amplitude
\beq
A = \int d\sz f(\sz) \bra{\{k^i_{rn}\}}   e^{\sum_{rs,imn} N^{rs}_{i,nm} a_{irn}^\dagger a_{ism}^\dagger  } \prod_{i/\ei=+1} \delta(p^i_1+p^i_2) \ket{0} ,
\eeq
 where the coefficients $N^{rs}_{i,nm}$ are the Fourier modes of the Green function as defined in the appendix and there is an undetermined 
measure $f(\sz)$. 
 The state $\ket{\{k^i_{rn}\}}$ represents the initial and final states in a occupation number notation. $k^i_{rn}$ is the occupation number of the $n$ 
oscillator mode in direction $i$ and for the string $r$, $r=1$ being the initial string and $r=2$ the final one. The derivation of this formula is
rather lengthy and can be found in Chapter 11 of \cite{GSW} as mentioned before. The relevant point here is that, having such formula we can check that the 
resulting   $ N^{rs}_{i,nm}$ that we compute in the appendix, satisfy all required equations (\ref{firstNeq})-(\ref{lastNeq}). What we obtain
is that $\NN$ can be written as (for $m+n\neq 0$):
\beq
 \NNa = -\frac{i}{8}\frac{(1+\ei)}{m+n} \left(a^{r}_m \delta_{n0} + a^s_n \delta_{m0} \right) 
  + \frac{1}{(m+n)\sin\sz} \mathrm{Im}\left(f^r_m f^s_n\right) .
\label{Nfactorized}
\eeq
 The coefficients $f^r_m$ and $a^r_m$ are given by linear combinations of Legendre polynomials as described in the appendix. 

  To find the measure we can compute the scattering of a tachyon from a D-brane. Since the tachyon is the vacuum state we get 
\beq
A = \bra{0} \int d\sz f(\sz)   e^{\sum_{rs,imn} N^{rs}_{i,00} a_{ir0}^\dagger a_{is0}^\dagger  } \prod_{i/\ei=+1} \delta(p^i_1+p^i_2) \ket{0}  .
\eeq
 From the appendix, the 00 component of the Neumann coefficients is:
\beq
N^{rs}_{i,00} = (1+\epsilon_i) \delta^{rs} \ln\left(\cos\frac{\sz}{2}\right) + \frac{1-\epsilon_i}{2}\ln\left(\sin\frac{\sz}{2}\right) .
\eeq
Which gives
\beq
A = \int d\sz f(\sz) \left[\cos\frac{\sz}{2}\right]^{4k^2} \left[\sin\frac{\sz}{2}\right]^{q^2} ,
\eeq
where $k$ is the momentum of the tachyon parallel to the brane (which is conserved) and $q=p_1+p_2$ is the momentum transfer.
 We also note that since $X^+=\s$ then $P^-=P_\s$ which vanishes for a closed string. Therefore we can only propagate states with
$P^-=0$. If we propagate a boundary state this is not a problem because all parallel momenta vanish for them. If we want
to extend the result for generic states we can rely on Lorentz invariance in the direction parallel to the brane and take $k^2$ to be the
full parallel momentum in the final amplitude. 

 Now, following~\cite{Klebanov:1995ni}, we compute the tachyon scattering amplitude using two vertex operators $V=e^{ipX}$ inserted on the 
half-plane. The boundary conditions determine the propagators to be 
$\langle X^\mu(z) X^\nu(w) \rangle = - \eta^{\mu\nu} \ln(z-w)$, $\langle X^\mu(z) \tilde{X}^\nu(\bar{w}) \rangle 
= - S^{\mu\nu} \ln(z-w)$, where $S^{\mu\nu}$ is equal to $\eta^{\mu\nu}$ up to a minus sign for Dirichlet coordinates. The amplitude is proportional to
\beqa
A &\sim& \int d^2z_1\, d^2z_2\ \langle V(z_1,\bar{z_1})V(z_2,\bar{z_2}) \rangle \\  
  &\sim&  \int d^2z_1\, d^2z_2\ |z_1-\bar{z}_1|^{p_1Dp_1} |z_1-z_2|^{2p_1p_2} |z_1-\bar{z}_2|^{2p_1Dp_2} |z_2-\bar{z}_2|^{p_2Dp_2} ,
\eeqa
where $p_iDp_j=p_{i\mu}p_{j\nu}S^{\mu\nu}$. The momenta satisfy $p_{1,2}^2=-m^2$, with $m$ the mass of the tachyon. The expression is $SL(2,\mathbb{R})$
invariant if  $m^2=-2$ which determine the tachyon mass. In that case we can factorize the volume of the Moebius group and set $z_1=i$, $z_2=iy$
getting (up to a normalization constant $c_3$): 
\beqa
A &=& 4c_3\int_0^1 dy (1-y^2)\, 2^{p_1Dp_1}\, (1-y)^{2p_1p_2}\, (1+y)^{2p_1Dp_2}\, (2y)^{p_2Dp_2} \\
  &=&  c_3 \int_0^{\pi} d\sz \left[\cos\frac{\sz}{2}\right]^{4k^2-3}\left[\sin\frac{\sz}{2}\right]^{q^2-3} ,
\eeqa
where we change variables $y=(1-\sin\frac{\sz}{2})/(1+\sin\frac{\sz}{2})$ as before and also decomposed the momenta into parallel 
$(p_1)_{\parallel}=-(p_2)_{\parallel}=k$ and perpendicular. The perpendicular one only entered as $q=p_1+p_2$.  Comparing with the previous expression we get that
$f(\sz)=8\left[\sin\sz\right]^{-3}$ as in \cite{Green:1994ix}.

 Therefore the operator $\Pp_0$, corresponding to a slit whose center is at $\sigma=0$, is given by
\beq
\Pp_0 = 8 c_3\int_0^\pi d\sz  \frac{1}{\sin^3\sz} e^{\sum_{rs,imn} N^{rs}_{i,nm} a_{irn}^\dagger a_{ism}^\dagger} \prod_{i/\ei=+1} \delta(p^i_1+p^i_2) \ket{0} ,
\label{Pop1}
\eeq
 where $\prod_{i/\ei=+1}$ is over the direction parallel to the brane (excluding the light-cone directions). We now have to integrate over the position of the 
center of the slit which gives
\beq
\Pp = \int_{0}^{2\pi}\!\! d\sigma\ T_{\sigma} \Pp_0 T^{-1}_\sigma ,
\label{TP}
\eeq
where $T_\sigma$ is the operator representing a translation in direction sigma by an amount $\sigma$. This gives 
\beq
\Pp = \frac{1}{\pi^3}\int_0^{2\pi}d\sigma 
    \int_0^\pi d\sz  \frac{1}{\sin^3\sz} e^{\sum_{rs,imn} N^{rs}_{i,nm} a_{irn}^\dagger a_{ism}^\dagger e^{-i(m+n)\sigma}} \prod_{i/\ei=+1} \delta(p^i_1+p^i_2) \ket{0} ,
\label{Pop2}
\eeq
where we anticipated that $c_3=(2\pi)^{-3}$ from the next subsection. 
 
 In principle we have established that the Hamiltonian $H=H_0-\lambda P$ with $\Pp$ given by the previous expression, sums the planar diagrams of a bosonic
string. We can have a problem however if there are singularities that need to be subtracted when two slits come together. If such extra terms are needed, they
modify the operator $\Pp$. It should be important to settle this issue. In any case we show in the next subsection that the operator $\Pp$ already contains
important information about the closed string background.

\subsection{Limit of small hole ($\sz\rightarrow 0$).}

 The operator $\Pp$ acts non-locally on the closed string. It is clear however that it contains a local part corresponding 
to very small holes. It is interesting to extract such part since it contains a pole in the transverse momentum transfer 
$q^2$ as $q^2\rightarrow 0$ and therefore can be described as a propagation of a string in a modified background. 

 Formally the limit corresponds to taking $\sz\rightarrow 0$ in $\Pp(\sz)$ which can be easily done using the properties of the
Neumann coefficients that we give in the appendix (see eq. (\ref{Nsz0})). Expanding the exponent in $\Pp_0$ (\ie eq.(\ref{Pop1})) at quadratic order we obtain:
\beqa      
 \sum_{rs,imn}\!\!\rule{0pt}{12pt}'\, N^{rs}_{imn} a_{irm}^\dagger a_{isn}^\dagger 
                       &=& \sum_{i, m\neq 0} \frac{1}{|m|} (a_{i1m}^{\dagger} +a_{i2m}^{\dagger} )(a_{i10}^{\dagger}+a_{i20}^{\dagger}) 
                                                     -\sum_{i, m\neq 0} \frac{1}{|m|} a_{i1m}^{\dagger}a_{i2,-m}^{\dagger} \nonumber \\
                       && +  \sum_{i, m\neq 0} \frac{1+\ei}{2}\sz(a_{i1m}^{\dagger}-a_{i2m}^{\dagger})(a_{i10}^{\dagger}+a_{i20}^{\dagger}) \nonumber \\
                       && -\frac{\sz^2}{8} \sum_{i, m\neq 0} (1+\ei)(a_{i1m}^{\dagger}-a_{i2m}^{\dagger})(a_{i10}^{\dagger}-a_{i20}^{\dagger}) \nonumber\\ 
                       && -\frac{\sz^2}{8} \sum_{i, m\neq 0} (1-\ei)|m| (a_{i1m}^{\dagger}+a_{i2m}^{\dagger})(a_{i10}^{\dagger}+a_{i20}^{\dagger}) \nonumber\\
                       && +\frac{\sz^2}{8} \sum_{i, m> 0,n>0}\left\{ -(a_{i1m}^{\dagger}-\ei a_{i2m}^{\dagger})(a_{i1n}^{\dagger}-\ei a_{i2n}^{\dagger}) \right. \nonumber \\
                       && \ \ \ \           -(a_{i1,-m}^{\dagger}-\ei a_{i2,-m}^{\dagger})(a_{i1,-n}^{\dagger}-\ei a_{i2,-n}^{\dagger})  \nonumber \\
                       && \ \ \ \    \left. -2\ei(a_{i1m}^{\dagger}-\ei a_{i2m}^{\dagger})(a_{i1,-n}^{\dagger}-\ei a_{i2,-n}^{\dagger}) \right\} ,
\eeqa
where the prime in the sum indicates that we extracted the term with $m=n=0$ that we consider separately. The sum over $i$ runs over all spatial directions
excluding the light-cone directions. 

 The first thing is to identify the zero modes with the momenta: $a_{ir0}^\dagger=p_{ir}$. Having done that we see that the term linear in $\sz$ is nonzero if 
$\ei=1$ which corresponds to Neumann boundary conditions. However, in that case the momentum is conserved, namely $p_{i1}+p_{i2}= a_{i10}^{\dagger}+a_{i20}^{\dagger}=0$.
This eliminates the linear term and the first correction is quadratic. 

The zero order term gives $\exp( -\sum_{i, m\neq 0} \frac{1}{|m|} a_{i1m}^{\dagger}a_{i2,-m}^{\dagger}) $ which should correspond to the identity operator, since,
for $\sz=0$ there is no
hole. To see that, consider just one mode $m$ and normalize canonically the creation operators by defining $\bar{a}_{irm}^\dagger= a_{irm}^\dagger/\sqrt{|m|}$. Then we get
\beq
 e^{ \bar{a}_{i1m}^\dagger \bar{a}_{i2,-m}^\dagger} \ket{0}_{1,m}\otimes \ket{0}_{2,-m} 
  = \sum_{N_m=0}^{\infty} (-1)^{N_m} \ket{N_m}_{1m} \otimes \ket{N_m}_{2,-m} = \ket{\Id}_m .
\eeq 
 The first thing we see is that this state, that we denote as $\ket{\Id}_m$ is the identity operator that identifies a state of string one with occupation
number $N_m$ with a state of the same occupation number on string two but on the mode $-m$. This is because a left moving excitation on string one looks like 
a right moving excitation when seen form the other end, namely string two. Moreover we can see that 
\beq  
 a_{i1m}^\dagger \ket{\Id}_m = -a_{i2,-m} \ket{\Id}_m .
\eeq
 This means that we can consider all operators as acting on string two. The resulting operator is a normal ordered function of $a_{i2m}^\dagger$ and
$a_{i2m}$ acting on $\ket{\Id}_m$. It is easy to see that such function is the standard representation of the operator in the oscillator basis, where
we can now eliminate the label $2$. In this way we go from the representation of the operator in terms of a two string vertex into the standard 
representation in terms of annihilation and creation operators. Using the rule that  $a_{i1m}^\dagger \rightarrow -a_{i2,-m} $ we get that
\beq
(a_{i1m}^{\dagger}-\ei a_{i2m}^{\dagger}) \rightarrow 
   \left\{\begin{array}{llcl}\ei=+1, \ \  &(a_{i1m}^{\dagger}-\ei a_{i2m}^{\dagger}) &=& -2p_{-m} \\  
                             \ei=-1,      &(a_{i1m}^{\dagger}-\ei a_{i2m}^{\dagger}) &=& i |m| x_{-m}\end{array} \right. ,
\eeq
where $x_m$ and $p_m$ are the Fourier components of position and momenta of string two. 
 Replacing in the previous expansion we obtain
\beq
 \sum_{rs,imn}\!\!\rule{0pt}{12pt}'\, N^{rs}_{imn} a_{irm}^\dagger a_{isn}^\dagger  \rightarrow iq\bar{y} + \frac{i}{4} \sz^2 q.y''-\frac{\sz^2}{8}y'y' +
 \sz^2 \bar{p}k - \frac{\sz^2}{2} \bar{p}\bar{p} +\ldots ,
\eeq
 where the operators are evaluated at $\sigma=0$ since we obtain the sums of Fourier modes without the exponents $e^{im\sigma}$. This is correct since the 
small hole is inserted at $\sigma=0$ so we expect $\Pp_0$ to act only there. On the other hand, the zero mode contribution ($m=n=0$) together with the measure $f(\sz)$ give
\beq
\frac{\left[\cos\frac{\sz}{2}\right]^{4k^2}\left[\sin\frac{\sz}{2}\right]^{q^2}}{\sin^3\sz}  \simeq \frac{\sz^{q^2-3}}{2^{q^2}} 
            \left[1-\sz^2\left(\half k^2+\frac{1}{24} q^2 - \half +\ldots\right)\right] .
\eeq
 Putting all together, we find,
\beq
\Pp_0 \simeq 2c_3  \left(\frac{\pi}{2}\right)^{q^2-2} \frac{1}{q^2-2} e^{iq\bar{y}} - \frac{8c_3}{q^2} \left(\frac{\pi}{2}\right)^{q^2} 
      \left(\half(k-\bar{p})^2 + \frac{1}{8}y'y' + \frac{1}{24} q^2 - \half - \frac{i}{4}qy''\right) e^{iq\bar{y}} .
\eeq
 The first term has a pole at the tachyon mass and therefore is an artifact of keeping only the bosonic sector. We now concentrate in the
$q^2\rightarrow 0$ pole and keep only the singular terms (\ie\ we put $q^2=0$ everywhere except in the denominator).
 The resulting operator is in a mixed base since the zero mode of $Y$ is in momentum representation and the other modes are in position representation.
We can change basis:
\beqa
 \bra{y_{20}} \Pp_0 \ket{y_{10}} &=& \int \frac{d^6q_1 d^6q_2}{(2\pi)^6} e^{iq_1y_{10}+iq_2y_{20}} \bra{q_2}\Pp\ket{q_1} \\
                             &=& \int \frac{d^6q_1 d^6q_2}{(2\pi)^6} e^{iq_1y_{10}+iq_2y_{20}} \Pp(q=q_1+q_2)        \\
                             &=& \delta^{(6)}(y_{10}-y_{20}) \int d^6 q e^{iqy_{20}} \Pp(q)  ,
\eeqa
where we used that the matrix element of $\Pp$ depends only on $q=q_1+q_2$. The delta function $\delta^{(6)}(y_{10}-y_{20})$ should be interpreted as
$\bra{y_{20}} y_{10}\rangle$ and therefore the rest is the operator in this basis. We now also use that
\beq
 \Pi_X = \frac{1}{2\pi} (-k+\bar{p}), \ \ \ \ \ Y = y_{20} + \bar{y} ,
\eeq
 since $k=p_1=-p_2$ is the zero mode part of the momentum and $\bar{p}$ is the oscillator part and the same with $y_{20}$ and $\bar{y}$.
 Finally we need
\beq
 \int d^6q \frac{1}{q^2} e^{iqy} = \frac{16\pi^3}{y^4}, \ \ \ \  \int d^6q \frac{q_j}{q^2} e^{iqy} = - \frac{64\pi^3}{y^6} y_j ,
\eeq
to obtain the result that
\beq
 \Pp_0 \simeq -\frac{16\pi^3c_3}{Y^4} \left((4\pi \Pi_X)^2 + Y'Y' - 4 + 4 \frac{YY''}{Y^2} \right), \ \ \ \ (\mbox{$q^2\rightarrow 0$ pole part}) ,
\eeq
 where all fields on the right hand side are evaluated at $\sigma=0$. One thing to notice is that, in the denominator, we obtained $Y$, the 
full quantum operator, not only the zero mode. To apply such operator we should remember that it is normally ordered. 

Now we have to translate an integrate on sigma (see eq.(\ref{TP}) and (\ref{Pop2}) ). The result is simply to 
evaluate the fields at $\sigma$ and integrate:
\beq
 \Pp \simeq -\int_{-\pi}^{\pi} d\sigma \frac{16\pi^3c_3}{Y^4} \left((4\pi\Pi_X)^2 + Y'Y' - 4 + 4 \frac{YY''}{Y^2} \right), \ \ \ \ (\mbox{$q^2\rightarrow 0$ pole part}) .
\eeq
 To this we should add the free Hamiltonian:
\beq
H_0 = \frac{1}{8\pi} \int (4\pi\Pi_X)^2 + (4\pi\Pi_Y)^2 + X'{}^2 + Y'{}^2 
\eeq
 Now we can compare with (\ref{Pbulk}). After identifying $\ap=2$, we see that there is partial agreement if we identify $c_3=(2\pi)^{-3}$. However, there are extra 
terms that can be attributed to the fact that we consider only the bosonic sector. It should be interesting to do the full superstring calculation to see if those terms
disappear. In the next subsection we do some preliminary steps in that direction but leave the full computation for future work. Of course there are also all the extra 
terms that do not have a pole in $q^2\rightarrow 0$. These terms we interpret as further corrections to the Hamiltonian. In fact, on the closed string side we
only computed the classical Hamiltonian whereas the one we compute here is supposed to be the full quantum Hamiltonian of the closed string. We should note that 
there might be further corrections if one has to subtract infinities when to slits come close together. It is known that this is a problem in the superstring. 
It should be important to understand this issue in more detail. It should also be interesting to study small holes in a covariant gauge perhaps following 
ideas in~\cite{Fischler:1987gz, Polyakov:2001af}.

\subsection{The superstring}

 In this section, for completeness we include the computation of the operator $\Pp$ in the superstring case. This is because the Neumann coefficients 
necessary for the superstring are the same as the one we already have. However, the superstring also has extra operators that should be included in the
end point of the slit where the Mandelstam map is singular. Although these operators are well understood, the consequences of including them in the
operator $\Pp$ are subtle, in particular if we investigate the $\sz\rightarrow 0$ limit. For that reason we leave this interesting problem for future
work and present here only the results that are essentially an extension of the calculations we have already done. Moreover, if we were to compute $\Pp(\sz\rightarrow 0)$
as we did in the bosonic sector, the action of the superstring moving in the full D3-brane background is not known so we cannot compare to anything. More interesting 
is to consider its field theory limit where the operator $\Pp$ contains the information of the sum of planar diagrams of the \N{4} SYM theory. In this 
case we can compare to the action~\cite{Metsaev:1998it} written in $\sigma$-gauge. However instead of taking the field theory limit of 
the stringy expression, it seems easier to derive $\Pp$ directly from the field theory using a slightly different method. We expect to report on this
in the near future.  

 For the superstring we consider $IIB$ string theory in the $SU(4)\times U(1)$ light-cone Green-Schwarz formalism, again following \cite{GSW} to which we refer the reader
for notation and conventions\footnote{This section is a bit aside of the main line of development so we did not try to make it self-contained.}. 
We introduce a set of right moving fermions and their canonical conjugates with mode expansions
\beqa
 \theta^A &=& \sum_{n=-\infty}^{\infty} Q_n^A e^{in\sigma} \\
 \lambda_A &=& \frac{1}{2\pi} \sum_{n=-\infty}^{\infty} Q_{nA} e^{in\sigma} ,
\eeqa
and commutation relations
\beq
\{Q_{nA},Q^B_m\} = \delta_{m+n} \delta_A^B .
\eeq
 The index $A$ is in the fundamental or anti-fundamental of $SU(4)$ depending if it is an upper or lower index. 
 We also introduce left moving fermions with mode expansions 
\beqa
 \tilde{\theta}^A &=& \sum_{n=-\infty}^{\infty} \tilde{Q}_n^A e^{in\sigma} \\
 \tilde{\lambda}_A &=& \frac{1}{2\pi} \sum_{n=-\infty}^{\infty}\tilde{Q} _{nA} e^{in\sigma} ,
\eeqa
and commutation relations
\beq
\{\tilde{Q}_{nA},\tilde{Q}^B_m\} = \delta_{m+n} \delta_A^B .
\eeq
We have a set of linearly realized supercharges:
\beq
 Q^+_{A} = Q_{0A} , \ \ Q^{+A} = Q_0^A, \ \ \tilde{Q}^+_{A} = \tilde{Q}_{0A}, \ \ \tilde{Q}^{+A} = \tilde{Q}^A_0 ,
\eeq
and a set of non-linearly realized:
\beqa
Q_{-A} &=& 2\sqrt{2} \int_{-\pi}^{\pi} \rho^I_{AB} \cA^I \theta^B + 8\pi \int_{-\pi}^\pi \cA^L \lambda_A \\
\tilde{Q}_{-A}    &=& 2\sqrt{2} \int_{-\pi}^{\pi} \rho^I_{AB} \tilde{\cA}^I \tilde{\theta}^B + 8\pi \int_{-\pi}^\pi \tilde{\cA}^L \tilde{\lambda}_A \\
Q_{-}^{A} &=& -4\sqrt{2}\pi \int_{-\pi}^{\pi} \rho^{IAB} \cA^I \lambda_B + 4 \int_{-\pi}^\pi \cA^R \theta^A \\
\tilde{Q}_{-}^{A} &=& -4\sqrt{2}\pi \int_{-\pi}^{\pi} \rho^{IAB} \tilde{\cA}^I \tilde{\lambda}_B + 4 \int_{-\pi}^\pi \tilde{\cA}^R \tilde{\theta}^A ,
\eeqa
where
\beq
\cA^I = P^I-\frac{1}{4\pi} \partial_\sigma Y^I, \ \ \ \tilde{\cA}^I = P^I+\frac{1}{4\pi} \partial_\sigma Y^I ,
\eeq
 and the same for $\cA^{R,L}$. They have the commutation relations
\beq
 \left[\cA(\sigma),\cA(\sigma')\right] = -\frac{i}{2\pi}\partial_\sigma \delta(\sigma-\sigma'), \ \ \ 
 \left[\tilde{\cA}(\sigma),\tilde{\cA}(\sigma')\right] = \frac{i}{2\pi}\partial_\sigma \delta(\sigma-\sigma'), \ \ \ 
 \left[\cA(\sigma),\tilde{\cA}(\sigma')\right] = 0 .
\eeq

It is useful to have a list of supersymmetry variations of the different fields:
\beq
 \begin{array}{lclclcl}
\rule{0pt}{18pt} \left[Q_{-A},\cA^I\right] &=& \frac{i\sqrt{2}}{\pi} \rho^I_{AB} \partial_\sigma \theta^B ,& \ \ \ &  
                     \left[\tilde{Q}_{-A},\tilde{\cA}^I\right] &=& -\frac{i\sqrt{2}}{\pi} \rho^I_{AB} \partial_\sigma \tilde{\theta}^B  ,\\  
\rule{0pt}{18pt} \left[Q_{-A},\cA^R\right] &=& 4i \partial_\sigma \lambda_A ,& \ \ \ &  
                     \left[\tilde{Q}_{-A},\tilde{\cA}^R\right] &=& -4i \partial_\sigma \tilde{\lambda}_A             ,\\  
\rule{0pt}{18pt} \left\{Q_{-A},\theta^B\right\} &=& 8\pi \delta_A^B \cA^L   ,& \ \ \ &  
                     \left\{\tilde{Q}_{-A},\tilde{\theta}^B\right\} &=& 8\pi \delta_A^B \tilde{\cA}^L                ,\\  
\rule{0pt}{18pt} \left\{Q_{-A},\lambda_B\right\} &=& 2\sqrt{2} \rho^I_{AB} \cA^I   ,& \ \ \ &  
                     \left\{\tilde{Q}_{-A},\tilde{\lambda}_B\right\} &=& 2\sqrt{2} \rho^I_{AB} \tilde{\cA}^I         ,\\  
\rule{0pt}{18pt} \left[Q_{-}^{A},\cA^I\right] &=& -  2 i\sqrt{2} \rho^{IAB} \partial_\sigma \lambda_B ,& \ \ \ &  
                     \left[\tilde{Q}_{-}^{A},\tilde{\cA}^I\right] &=& 2 i\sqrt{2} \rho^{IAB} \partial_\sigma \tilde{\lambda}_B  ,\\
\rule{0pt}{18pt} \left[Q_{-}^{A},\cA^L\right] &=& \frac{2i}{\pi} \partial_\sigma \theta^A ,& \ \ \ &  
                     \left[\tilde{Q}_{-}^{A},\tilde{\cA}^L\right] &=& -\frac{2i}{\pi} \partial_\sigma \tilde{\theta}^A       ,\\  
\rule{0pt}{18pt} \left\{Q_{-}^{A},\theta^B\right\} &=& -4\sqrt{2}\pi \rho^{IAB} \cA^I   ,& \ \ \ &  
                     \left\{\tilde{Q}_{-}^{A},\tilde{\theta}^B\right\} &=& -4\sqrt{2}\pi \rho^{IAB} \tilde{\cA}^I            ,\\  
\rule{0pt}{18pt} \left\{Q_{-}^{A},\lambda_B\right\} &=& 4 \delta^A_B \cA^R   ,& \ \ \ &  
                     \left\{\tilde{Q}_{-}^{A},\lambda_B\right\} &=& 4 \delta^A_B \tilde{\cA}^R                               ,\\
\rule{0pt}{28pt}\left\{Q^A_+,\lambda_B\right\} &=& \frac{1}{2\pi} \delta^A_B  &&             
                     \left\{\tilde{Q}^A_+,\tilde{\lambda}_B\right\} &=& \frac{1}{2\pi} \delta^A_B    ,\\             
\rule{0pt}{18pt}\left\{Q_{+A},\theta^B\right\} &=& \delta_A^B  &&             
                     \left\{\tilde{Q}_{+A},\tilde{\theta}^B\right\} &=& \delta_A^B  
 \end{array}
\eeq             
 The supersymmetry algebra is
\beqa
\rule{0pt}{18pt} \left\{Q^+_A,  Q^-_B\right\} &=& \sqrt{2} P^I \rho^I_{BA} ,\\
\rule{0pt}{18pt} \left\{Q^+_A,  Q^{-B}\right\} &=& 2 P^R \delta_A^B ,\\
\rule{0pt}{18pt} \left\{Q^{+A}, Q^-_B\right\} &=& 2 P^L \delta^A_B ,\\
\rule{0pt}{18pt} \left\{Q^{+A}, Q^{-B}\right\} &=& -\sqrt{2} P^I \rho^{IBA} ,\\
\rule{0pt}{18pt} \left\{Q^{-A}, Q^-_B\right\} &=& 2 H \delta^A_B ,\\
\rule{0pt}{18pt} \left\{Q^{+A}, Q^+_B\right\} &=& \delta^A_B ,
\eeqa
 and the same in the left moving sector. The Hamiltonian $H$ is given by
\beq
 H = P^2 + 4N+4\tilde{N} ,
\eeq
 where $N$, $\tilde{N}$ are the left and right moving occupation numbers ($N=\tilde{N}$ by the level matching condition).
 In terms of oscillators they are given by
\beqa
 N &=& \sum_{n\ge 1} (\alpha^i_n)^\dagger \alpha^i_n + \sum_{m\ge 1} m Q^B_{-m} Q_{Bm} + \sum_{m\ge 1} m Q_{B,-m}Q^B_m ,\\
\tilde{N} &=& \sum_{n\ge 1} (\tilde{\alpha}^i_n)^\dagger \tilde{\alpha}^i_n 
                   + \sum_{m\ge 1} m \tilde{Q}^B_{-m} \tilde{Q}_{Bm} + \sum_{m\ge 1} m  \tilde{Q}_{B,-m}\tilde{Q}^B_m ,,
\eeqa
 The occupation numbers are positive so we define the vacuum $\ket{0}$ as
\beq
Q_{mB} \ket{0}=0, \ \ Q_{m}^{B} \ket{0}=0, \ \ \alpha^i_n\ket{0}=0 , \ \ \ (m\ge 1) ,
\eeq
and the same for the left movers.

 Now we have to find out what condition we should impose on the boundary state $\ket{B}$ that corresponds to the D3-brane. 
We can try the following conditions compatible with the $SU(4)\times U(1)$ symmetry. 
\beqa
 \left( \cA^{L,R} + \varepsilon_{L,R} \tilde{\cA}^{L,R} \right)  \ket{B} &=& 0 , \\ 
 \left( \cA^I+\varepsilon_\perp \tilde{\cA}^I \right)  \ket{B}&=& 0   ,\\
 \left( \theta^A - \mu \tilde{\theta}^A \right)  \ket{B} &=&0 , \label{msg}\\
 \left( \lambda_A + \frac{1}{\mu}\tilde{\lambda}_A \right)  \ket{B} &=& 0 .
\eeqa   
 Since these conditions are imposed on a state, they have to commute among themselves. This implies the relation 
between the last two conditions for $\theta$ and $\lambda$. In the open string channel one has to impose conditions on the coordinates and
therefore they have to be compatible with the canonical commutation relations (in that case one does not get the minus sign in (\ref{msg})). 
 The conditions should preserve half the supersymmetries. As candidates to preserved supersymmetries we take
\beq
Q_+^A + \nu_1 \tilde{Q}^A_+, \ \ Q_{+A} + \nu_2 \tilde{Q}_{+A}, \ \ Q_{-A}+\nu_3\tilde{Q}_{-A}, \ \ Q_{-}^A + \nu_4 \tilde{Q}_{-}^A .
\eeq
 These supersymmetries should commute with all the previous conditions. We then get that
\beq
 \varepsilon_L=\varepsilon_R, \mu^2 = -\varepsilon_\perp\varepsilon_L, \nu_1=-\mu, \ \nu_2=\frac{1}{\mu}, 
 \ \ \nu_3 = \varepsilon_\perp \mu, \nu_4 = \varepsilon_R \mu,
\ \varepsilon_L^2=1, \varepsilon_\perp^2=1 ,
\eeq
 which have the solutions
\begin{itemize}
\item{D1 brane}, $\varepsilon_L=\varepsilon_\perp=-1$, \ $\mu=\pm i$,\ $\nu_3=\mp i$,\ $\nu_4=\mp i$ ,
\item{D3 brane}, $\varepsilon_L=1$, $\varepsilon_\perp=-1$,\ $\mu=\pm 1$,\ $\nu_3 = \mp 1$,\ $\nu_4=\pm 1$ ,
\item{D9 brane}, $\varepsilon_L=\varepsilon_\perp=1$, \ $\mu=\pm i$,\ $\nu_3=\pm i$,\ $\nu_4=\pm i$ .
\end{itemize}
 The two signs correspond to branes and anti-branes.
 We are interested in D3-branes so we impose
\beqa
 \left( \cA^{L,R} + \tilde{\cA}^{L,R} \right)  \ket{B} &=& 0 , \\ 
 \left( \cA^I- \tilde{\cA}^I \right)  \ket{B}&=& 0   \\
 \left( \theta^A  - \tilde{\theta}^A \right)  \ket{B} &=&0 , \\
 \left( \lambda_A + \tilde{\lambda}_A \right)  \ket{B} &=& 0 ,
\eeqa   
 which preserve
\beq
Q_+^A - \tilde{Q}^A_+, \ \ Q_{+A} + \tilde{Q}_{+A}, \ \ Q_{-A}-\tilde{Q}_{-A}, \ \ Q_{-}^A +  \tilde{Q}_{-}^A .
\eeq
This is regarding a boundary state. In the case of the vertex $\ket{V}$ we should impose these conditions on the slit and continuity of
the coordinates in the rest. This leads to the conditions
\beqa
 \left(\theta_1^A   - \theta_2^A  - \tilde{\theta}_1^A  + \tilde{\theta}_2^A  \right) \ket{V} &=& 0 ,\ \ \ -\pi\le\sigma\le \pi ,\\   
 \left(\lambda_{1A}  + \lambda_{2A} + \tilde{\lambda}_{1A} + \tilde{\lambda}_{2A} \right) \ket{V} &=& 0 ,\ \ \ -\pi\le\sigma\le \pi ,\\  
 \left(\theta_1^A   - \theta_2^A  + \tilde{\theta}_1^A  - \tilde{\theta}_2^A  \right) \ket{V} &=& 0 ,\ \ \ \sz\le|\sigma|\le \pi ,\\  
 \left(\lambda_{1A}  + \lambda_{2A} - \tilde{\lambda}_{1A} - \tilde{\lambda}_{2A} \right) \ket{V} &=& 0 ,\ \ \ \sz\le|\sigma|\le \pi ,\\  
 \left(\theta_1^A   + \theta_2^A  - \tilde{\theta}_1^A  - \tilde{\theta}_2^A  \right) \ket{V} &=& 0 ,\ \ \ -\sz\le\sigma\le \sz ,\\  
 \left(\lambda_{1A}  - \lambda_{2A} + \tilde{\lambda}_{1A} - \tilde{\lambda}_{2A} \right) \ket{V} &=& 0 ,\ \ \ -\sz\le\sigma\le \sz . 
\eeqa
 To construct the vertex state it is useful to define new fermionic variables:
\beq
\begin{array}{lclclcl}
\rule{0pt}{18pt} \Xi^A  &=& \frac{1}{\sqrt{2}} \left(\theta_1^A+\tilde{\theta}_2^A\right)     , &\ \ & \bar{\Xi}_A  &=& \frac{1}{\sqrt{2}} \left( \lambda_{1A} + \tilde{\lambda}_{2A} \right) ,\\
\rule{0pt}{18pt} \chi_A &=& \frac{1}{\sqrt{2}} \left(\lambda_{2A}+\tilde{\lambda}_{1A}\right)  , &&     \bar{\chi}^A &=& \frac{1}{\sqrt{2}} \left( \theta_2^A  + \tilde{\theta}_1^A  \right) ,\\
\rule{0pt}{18pt} c_A    &=& \frac{1}{\sqrt{2}} \left(\tilde{\lambda}_{1A}- \lambda_{2A} \right), &&     \bar{c}^A    &=& \frac{1}{\sqrt{2}} \left( \tilde{\theta}_1^A - \theta_2^A     \right) ,\\
\rule{0pt}{18pt} d^A    &=& \frac{1}{\sqrt{2}} \left(\theta_1^A - \tilde{\theta}_2^A\right)  , &&     \bar{d}_A    &=& \frac{1}{\sqrt{2}} \left( \lambda_{1A} - \tilde{\lambda}_{2A} \right) 
\end{array}
\eeq
 in terms of which the conditions are
\beqa
 \left(\chi_A  + \bar{\Xi}_A  \right) \ket{V} &=& 0 ,\ \ \ -\pi\le\sigma\le \pi   ,\\   
 \left(\Xi^A   - \bar{\chi}^A \right) \ket{V} &=& 0 ,\ \ \ -\pi\le\sigma\le \pi   ,\\  
 \left(\bar{c}^A +      d^A   \right) \ket{V} &=& 0 ,\ \ \  \sz\le|\sigma|\le \pi ,\\  
 \left(\bar{d}_A -      c_A   \right) \ket{V} &=& 0 ,\ \ \  \sz\le|\sigma|\le \pi ,\\  
 \left(     d^A  - \bar{c}^A  \right) \ket{V} &=& 0 ,\ \ \ -\sz\le\sigma\le \sz   ,\\  
 \left(\bar{d}_A +      c_A   \right) \ket{V} &=& 0 ,\ \ \ -\sz\le\sigma\le \sz .
\eeqa
The first condition is solved by the state
\beq
 e^{\sum_{m\ge 1} \left(\chi_{mA} \Xi^{A}_{-m} + \bar{\Xi}_{A,-m} \bar{\chi}^A_{m}\right) } \prod_B (\chi_{0B}+\bar{\Xi}_{0B}) \ket{0} .
\label{stf1}
\eeq
The other four conditions require more work. We introduce yet another set of fermionic modes
\beq
\begin{array}{lclcclcll}
 a_{nA}^\dagger &=& c_{nA}          &\ , \mbox{if}\ (n>0)  &\ \ \ & b_n^{A\dagger} &=&\bar{c}_n^A  &\ , \mbox{if}\ (n>0)    ,\\
 b_n^{A\dagger} &=& d_{n}^{A}       &\ , \mbox{if}\ (n<0)  &\ \ \ & a_{nA}^\dagger &=&\bar{d}_{nA} &\ , \mbox{if}\ (n<0)    ,\\
 a^A_n          &=& \bar{c}_{n}^{A} &\ , \mbox{if}\ (n<0)  &\ \ \ & b_{nA}         &=&{c}_{nA}     &\ , \mbox{if}\ (n<0)    ,\\
 b_{nA}         &=& \bar{d}_{nA}    &\ , \mbox{if}\ (n>0)  &\ \ \ & a_{n}^{A}      &=&{d}_{n}^{A}  &\ , \mbox{if}\ (n>0)    ,
\end{array}
\eeq  
 and propose a state
\beq
 \ket{V} = e^{\sum_{m,n\neq 0} V_{nm} |m| b^{A\dagger}_{n} a^{\dagger}_{mA} +\sum_{m\neq 0} (\bar{b}^A_{0} \alpha_m + \bar{a}^A_{0} \beta_m) a^{\dagger}_{mA}} \ket{0} ,
\label{stf2}
\eeq
which requires that
\beq
 |\sigma| \le\sz 
\left\{  \begin{array}{rcll}
 \sum_{n\neq 0} V_{nm} |m| e^{in\sigma} + 2\alpha_m &=& -e^{-im\sigma}, &\ \ (m\neq 0) ,\\
 \sum_{n\neq 0} n V_{mn} e^{in\sigma} &=& \sign(m) e^{-im\sigma}, &\ \ (m\neq 0) ,\\
 \sum_{m\neq 0} \sign(m) \alpha_m e^{-im\sigma} &=& 0 &,\\
 \sum_{m\neq 0} \sign(m) \beta_m e^{-im\sigma} &=& -1 & ,
\end{array} \right.
\eeq
and
\beq
 \sz \le |\sigma| \le \pi  
\left\{  \begin{array}{rcll}
 -\sum_{n\neq 0} \sign(n) V_{nm} |m| e^{in\sigma} + 2\beta_m &=& \sign(m) e^{-im\sigma},& \ \ (m\neq 0) ,\\
 \sum_{n\neq 0} |n|  V_{mn} e^{in\sigma} &=& e^{-im\sigma},& \ \ (m\neq 0) ,\\
 \sum_{m} \alpha_m e^{im\sigma} &=& 1 &,\\
 \sum_{m} \beta_m  e^{im\sigma} &=& 0 & .
\end{array} \right.
\eeq
 The zero modes were defined as 
\beqa
    {a}_{0A}   &=&\bar{d}_{0A}   -     c_{0A},       \\
\bar{a}_{0}^{A}&=&    {d}_{0}^{A}-\bar{c}_{0}^{A}, \\
    {b}_{0A}   &=&     c_{0A}    +\bar{d}_{0A},   \\
\bar{b}_{0}^{A}&=&\bar{c}_{0}^{A}+    {d}_{0}^{A},
\eeqa
 and obey
\beq
 \{a_{0A},\bar{a}_0^B\} = 2 \delta_A^B, \ \ \  \{b_{0A},\bar{b}_0^B\} = 2 \delta_A^B .
\eeq
 The vacuum obeys
\beq
 a_{0A} \ket{0} =0, \ \ \ \ b_{0A}\ket{0} =0 .
\eeq
 As usual, from the commutations relations we see that zero modes can be represented as gamma matrices.
 
 Going back to our main problem, using the properties of the Neumann coefficients that we derive in the appendix we obtain that 
the equations for the coefficients $V_{mn}$, $\alpha_m$, $\beta_m$ are solved by
\beqa
 V_{nm} &=& -2\left(\NC^{11}_{nm}(\ei=-1) + \NC^{12}_{nm}(\ei=-1)\right) ,\\
 \alpha_m &=& - |m| \left(\NC^{11}_{0m}(\ei=-1) + \NC^{12}_{0m}(\ei=-1)\right) ,\\
 \beta_m &=& - m \left(\NC^{12}_{m0}(\ei=1) - \NC^{11}_{m0}(\ei=1)\right) .
\eeqa
 This completely determines the state that imposes continuity and the correct conditions on the slit as the product of (\ref{stf1}) and (\ref{stf2}). 
However it is well-known that to reproduce the correct string amplitudes, one has to insert extra operators associated with the end points of 
the slit where the conformal map is singular. These operators are known in the open string sector. One just has to write them after doing 
a $\sigma\leftrightarrow \tau$ interchange. After that one should compute the measure by comparison with known D3-brane scattering amplitudes. 
We leave this problem for the future. 

 One comment that we have to make, however, is about terms of higher order in $\lambda$. The theory is still supersymmetric so we should now have that 
\beq
\left\{Q^{-A}, Q^-_B\right\} = 2 (H_0-\lambda \Pp) \delta^A_B ,\\
\eeq   
 This means that $Q^{-A}$ also has terms of order $\lambda$ at least: $Q^{-A}=Q^{-A}_0+\lambda Q^{-A}_1$. If $\{Q^{-A}_1,Q^-_{B1}\}\neq 0$ this implies
that $H$ has at least terms of order $\lambda^2$. If such higher order terms are present one might be able to determine them by the closure of the supersymmetry 
algebra. It is not obvious that this can be done since, already from the closed string point of view the theory is possibly not well-defined due to these
higher order terms that can appear in the Hamiltonian\footnote{I am grateful to N. Berkovits for a comment on this issue.}. A natural hope is that these problems
are absent if one wants to study just the field theory limit, what is, after all, the most important case. 

 About the limit $\ap\rightarrow 0$, it is possibly that is trivial in the case of the vertex. This is because the vertex has a naive scale invariance in 
$X\rightarrow \mu X$ and $Y\rightarrow Y/\mu$ since the boundary conditions are invariant under such rescaling. If it turns out that the naive invariance 
is valid at the quantum level, namely for the full vertex, then one should be able to rescale out $\ap$. In that case the only thing we would need to do 
is to modify $H_0$ as we found in section \ref{bkg} from the point of view of the closed string in the D-brane background.

\section{Relation to field theory and $\sigma\leftrightarrow\tau$ duals}
\label{ftsection}

 One would like to understand the planar diagrams directly in field theory. Although we do not analyze this problem in the present paper, there are 
some points which already follow from our previous discussion.

Any field theory with fields in the adjoint can be represented
in terms of diagrams as in fig.\ref{fig1} in light cone frame. The property however that we need is that the contribution of any such diagram can 
be represented also as a propagation in $\tau$ after a $\sigma\leftrightarrow\tau$ interchange. It appears from our discussion that this should be
possible whenever we can embed the theory in a string theory which has a local world-sheet action. In other cases it is not clear. However we 
should note that we are asking, not that the theory has a string dual but simply that it can be written as propagation of a Hamiltonian of the 
type $H_0-\lambda \Pp$ in the crossed channel. This appears a weaker condition and might be more easily satisfied. To be more precise, in the 
standard representation, the ``world-sheet'' theory describing the field theory is local in $\tau$ but not in $\sigma$. We want to see
if it is possible a representation local in $\sigma$ and possibly not in $\tau$. After a $\sigma\leftrightarrow\tau$ transposition the locality 
in $\sigma$ becomes local in $\tau$ which means that we have a Hamiltonian representation (with $H=H_0-\lambda\Pp$). This new $H$ we define
as the $\sigma\leftrightarrow\tau$ dual of the original theory which contains the information about the planar diagrams.

 Going back to section \ref{bkg}, the near-horizon limit suggested that 
\beq
H_0=\half \int d\sigma \left(\Pi_Y^2 +X'{}^2 \right)
\label{H0ft}
\eeq
should be the $\sigma\leftrightarrow\tau$ dual of a free field theory.
 Let us study what this Hamiltonian gives when used in a world-sheet diagram.
Consider first
\beq
 H_{01} = \half \int d\sigma X'{}^2
\eeq
The Hamiltonian is diagonal in the $X$ basis, therefore the propagator is
\beq
 \bra{X_f(\sigma)} e^{-\tau H_{01}} \ket{X_i(\sigma)} = \prod_\sigma \delta\left(X_f(\sigma)-X_i(\sigma)\right) e^{-\half\tau\int d\sigma X_i'{}^2}
\eeq
 In a diagram as that in figure \ref{figft1}(a) we have that the boundary conditions are Neumann, which is equivalent to say that we should integrate over 
all boundary values of $X$ with unit measure. In particular note that the values of $X$ are different on both sides of the slit. On the other
hand we should take into account that the propagator has a delta function which leads to the identification of boundary conditions as indicated. 
This gives for the value of such diagram
\beq
 Z_1 = \int \cD X_a(\s)\, \cD X_b(\s)\, \cD X_c(\s)  e^{-\half\tau\int_0^{\sigma_1}X_c'{}^2 
                                            -\half(\tau-\tau_0)\int_{\sigma_1}^{\sigma_1+\sigma_2} X_a'{}^2
                                            -\half \tau_0\int_{\sigma_1}^{\sigma_1+\sigma_2} X_b'{}^2 }
\eeq
 These are one-dimensional path integrals. Continuity of the coordinates and periodicity in sigma imply that
\beq
 X_a(\sigma_1) = X_b(\sigma_1) = X_c(\sigma_1) = x_1, \ \ \ \ X_a(0) = X_b(0) = X_c(0) = x_0
\eeq
 Therefore we get
\beqa
 Z_1 &=& \cN \int d^2x_1 d^2x_0 e^{-\frac{\tau}{2\sigma_1}(x_1-x_0)^2-\frac{\tau-\tau_0}{2\sigma_2}(x_1-x_0)^2-\frac{\tau_0}{2\sigma_2}(x_1-x_0)^2} \\
     &=& \cN \int d^2X_0 \int d^2x e^{-\frac{\tau}{2\sigma_1}x^2-\frac{\tau-\tau_0}{2\sigma_2}x^2-\frac{\tau_0}{2\sigma_2}x^2}
\eeqa
where we get a divergence from the zero mode $X_0=x_1+x_0$. It is clearly present since the Hamiltonian is proportional to $X'{}^2$ and 
seems to play no role so we discard it. More importantly we get also an infinite measure $\cN$ which depends on the variables 
$\tau$,$\tau_0$,$\sigma_1$ and $\sigma_2$. We do not attempt to evaluate it in this paper. 

 Consider now
\beq
 H_{02} = \half \int d\sigma \Pi_Y^2
\eeq
 The propagator is 
\beq
 \bra{Y_f(\sigma)} e^{-\tau H_{01}} \ket{Y_i(\sigma)} = \cN_2 e^{-\frac{1}{2\tau}\int d\sigma \left(Y_f(\sigma)-Y_i(\sigma)\right)^2}
\eeq
 again up to a normalization factor $\cN_2$. For the same diagram (see fig.\ref{figft1}(b))we now get
\beq
 Z_2 = \cN_2 e^{-\frac{\sigma_2 m^2}{2\tau}}e^{-\frac{\sigma_2 m^2}{2\tau_0}}
\eeq

\FIGURE{\epsfig{file=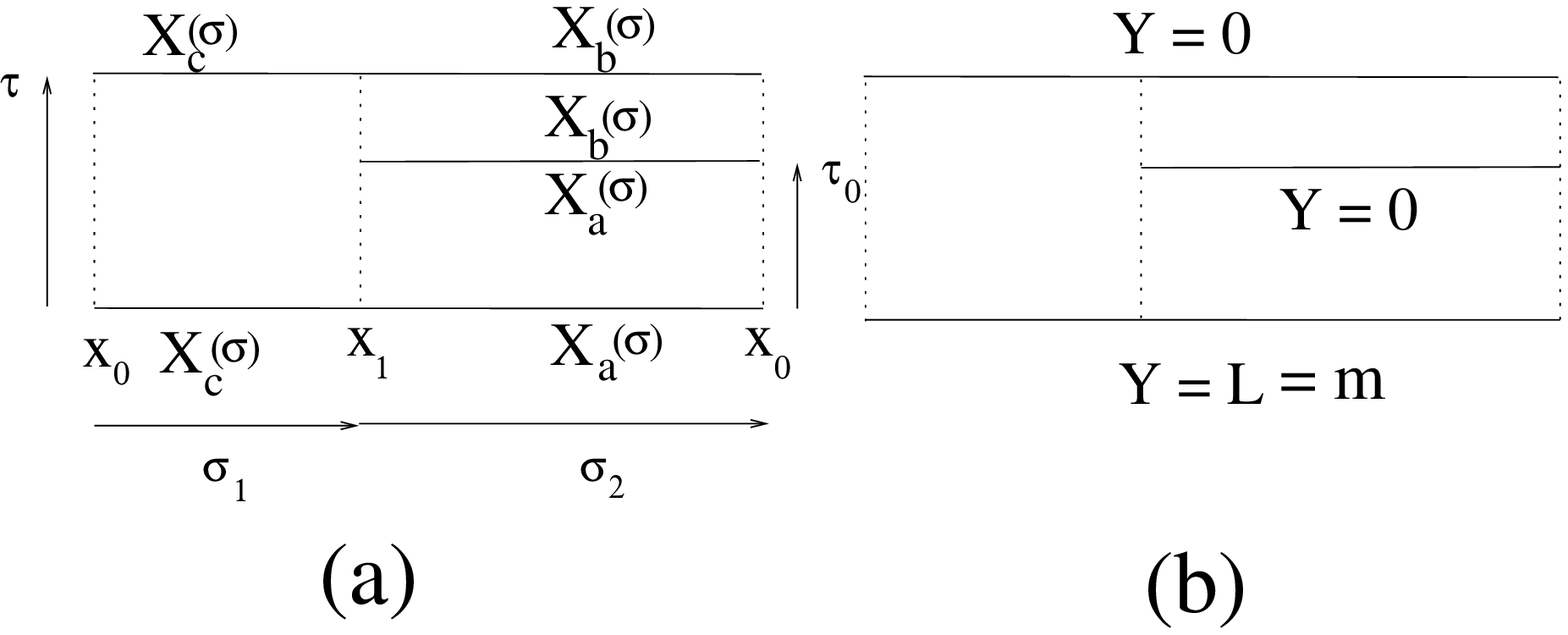, width=14cm}
\caption{One loop diagram computed with the Hamiltonian of eq.(\ref{H0ft}). The coordinate $X$ does not change under propagation but we should integrate over
all possible boundary conditions. The coordinate $Y$ is fixed at the boundary but it can change in time.}
\label{figft1}
}

 If we now make a $\sigma\leftrightarrow\tau$ rotation we get a diagram as in fig.\ref{figft2}. In light cone frame for a $\phi^3$ theory this diagram gives
\beq
Z = \int d^2p_\perp \int d^2k_{\perp} \frac{1}{|p^+|}e^{-\frac{p_{\perp}^2+m^2}{2p^+}t_0}
    \frac{1}{|k^+|}e^{-\frac{k_{\perp}^2+m^2}{2k^+}(t-t_0)}
    \frac{1}{|p^+-k^+|}e^{-\frac{(p_{\perp}-k_\perp)^2+m^2}{2(p^+-k^+)}(t-t_0)}
\eeq
Now we make the identifications
\beq
k^+=\tau_0, \ \ p^+=\tau, \ \ t_0=\sigma_1,\ \ \ t=\sigma_1+\sigma_2 ,
\eeq
and use the formula
\beq
 \int d^2x\ e^{-iq_\perp x} e^{-\half\frac{\tau x^2}{\sigma}} = \frac{2\pi\sigma}{\tau} e^{-\half \frac{\sigma q_\perp^2}{\tau}} . 
\eeq
We obtain
\beqa
 Z &=& \int \frac{d^2p_\perp d^2k_{\perp} d^2x_1 d^2x_2 d^2x_3}{(2\pi)^3\sigma_1\sigma_2^2} e^{-ip_\perp x_1+i k_\perp x_2+i(p-k)_\perp x_3} 
                               e^{-\frac{\tau x_1^2}{2\sigma_1}-\frac{\tau x_2^2}{2\sigma_2}-\frac{(\tau-\tau_0) x_3^2}{2\sigma_2}}
                               e^{-\frac{\sigma_1 m^2}{2\tau}-\frac{\sigma_2 m^2}{2\tau_0}} \nonumber\\
  &=&  \frac{1}{2\pi} \int d^2x  \frac{1}{\sigma_1\sigma_2^2} 
                               e^{-\frac{\tau x^2}{2\sigma_1}-\frac{\tau x^2}{2\sigma_2}-\frac{(\tau-\tau_0) x^2}{2\sigma_2}}
                               e^{-\frac{\sigma_1 m^2}{2\tau}-\frac{\sigma_2 m^2}{2\tau_0}}  . 
\eeqa
 In this way we recover a similar expression as one gets by multiplying $Z_1Z_2$ in the previous case. Of course nothing of this is truly meaningful if 
we do not find a way to regularize the measures $\cN_1$ and $\cN_2$ such that $Z=Z_1 Z_2$. We do not attempt to do so here. The correct measure
can only be obtained after considering all the fields in the theory. We only wanted to show that the Hamiltonian $H_0$ seems a reasonable
starting point to describe a field theory.  
 On the other hand, the fact that we obtained the correct dependence in $x$ is meaningful, even if it is integrated, because the position $x$ has the same
meaning on both calculations.  Namely, is the difference between the position of the particle at $\sigma=0$ and $\sigma=\sigma_1$ (or $t=0$ and $t=t_0$).

\FIGURE{\epsfig{file=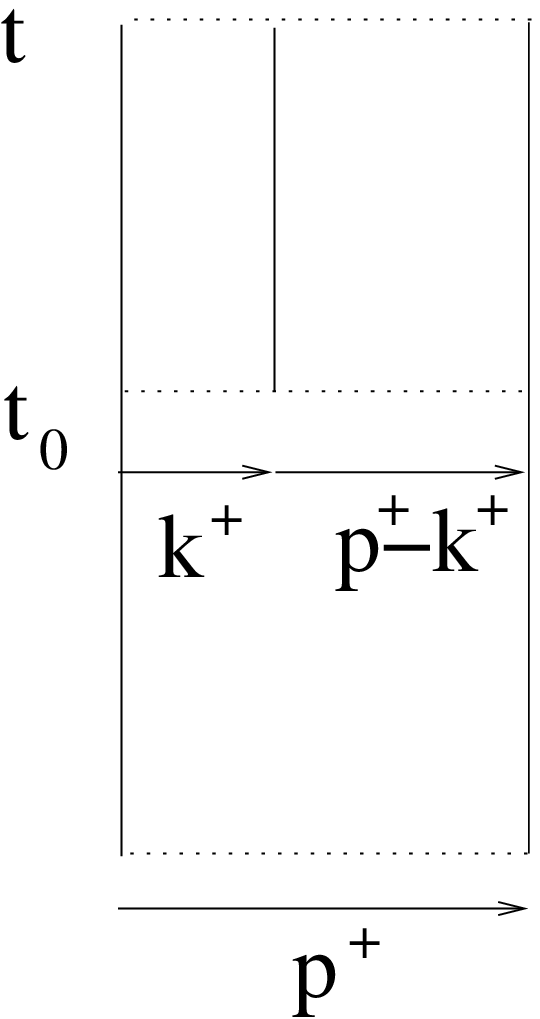, height=6cm}
\caption{Diagram of fig.\ref{figft1} after a $\sigma\leftrightarrow\tau$ rotation. It should be interpreted as a field theory diagram.}
\label{figft2}
}

\section{Conclusions}
\label{conclu}

 We have studied the sum of planar diagrams in an open string theory and concluded that, under certain assumptions about the general form of the
higher order diagrams, the sum exponentiates and determines a Hamiltonian of the form $H=H_0-\lambda \Pp$. We also discussed the possible existence
of higher order corrections that we were unable to determine at this stage.  On the other hand, this Hamiltonian should describe the
propagation of a closed string in a modified background which, in the same gauge, turned out to be linear in $\lambda$. In fact, in the limit of
small holes, we obtain partial agreement between the two pictures. The difference is attributed to the fact that we studied only the bosonic 
sector of the theory. 
 For the superstring a similar result should follow. However, in that case, it is even more clear that there could be also corrections of higher order 
in $\lambda$,  due to contact terms. It seems important to clarify this issue. In any case it seems to us that a great deal of information, 
\eg\ the form of the dual background, should be already contained in $\Pp$. The linear dependence in $\lambda$ is valid for any Dp-brane 
except that in the general case there is a position dependent dilaton in the background. It should be interesting to understand it
from the open string point of view. It should also be interesting to understand the operator $\Pp$ for D-brane bound states, moving branes etc. We should also note that the same ideas can be applied to the computation of scattering amplitudes. In this case the ``dual'' string
would be infinitely long (see fig.\ref{fig2}).   

 One can also think of applying these ideas directly to a field theory. What we need is that the planar diagrams in light cone gauge
have a dual interpretation as propagation in time in the ($\sigma\leftrightarrow\tau$) dual channel. This seems true if the field theory
can be embedded into a string theory. However one is asking less, so it can be a more generic property. Whenever this happens, and if the 
relative weight of the diagrams is correct, the sum of planar diagrams should exponentiate allowing one to define the 
$(\sigma\leftrightarrow\tau)$ dual Hamiltonian of the theory. This is a Hamiltonian for a one dimensional system and its properties
determine the properties of the planar diagrams. If this actually happens or not should be analyzed in each particular field theory. 
 
 A different avenue of investigation could be to sew two diagrams of the type appearing in fig.\ref{fig2} to construct a closed string amplitude. 
The corresponding operator $\Pp$ should be described by a four string vertex. The meaning of summing those diagrams however is not clear to us.

 To summarize, we have found that in this kind of light-cone gauge, that we call $\sigma$-gauge, the Hamiltonian of the string in a D-brane 
background is linear in $\lambda=g_sN$. We were able to reproduce this from the open string point of view with a naive exponentiation of the 
operator $\Pp$. This could be just a coincidence due to supersymmetry or it could mean something deeper. It remains to be seen what extra 
contributions, if any, the operator $\Pp$ has. It seems important also to understand if the operator $\Pp$ can be constructed directly in the
field theory. If it can be done, studying the operator $H=H_0-\lambda \Pp$ should give valuable information about the planar diagrams, for example,
a tachyonic ground state can be a signal of a phase transition such as confinement. As can be seen from eq.(\ref{ftlimbulk}), one thing that should 
follow in the field theory limit, is that whereas $H_0$ describes a ``fluffy'' or ``rigid'' string (depending on the direction), the operator $\Pp$ 
gives the string a tension.  We leave all these questions for future work.

%




\section{Acknowledgments}

I am grateful to N. Berkovits and Lars Brink for discussions on light-cone gauge and for pointing out some references. I am also 
indebted to I. Klebanov, J. Maldacena and A. Polyakov for comments on the initial version of this paper. 

This material is based upon work supported by the National Science Foundation Grant No. PHY-0243680. Any opinions, findings, and conclusions 
or recommendations expressed in this material are those of the authors and do not necessarily reflect the views of the National Science Foundation.

\appendix

\section{Neumann coefficients}
\label{Nmncalc}

 In this section we compute the Neumann coefficients. To do that we have to compute the Green 
function in the infinite cylinder with a slit where the field obeys Neumann or Dirichlet boundary 
conditions. One way to do that is to use a conformal map to map the cylinder with a slit to the upper 
half plane in such a way that the slit is mapped to the real axis.
 By composing the standard exponential map with appropriate inversions and translations such conformal 
transformation is not difficult to find. The result is
\beq
 \rho = \ln \frac{(z-i)(z+iy)}{(z+i)(z-iy)} ,
\eeq
where $\rho = \tau+i\s$ and $z$ parameterized the upper half plane. We illustrate this map in figure \ref{fig:map}.
The constant $y$ is related to the size of the slit by 
\beq
y = \frac{1-\sin\frac{\sz}{2}}{1+\sin\frac{\sz}{2}} .
\label{ydef}
\eeq
 The point $\tau=-\infty$ maps to $z=i$ and $\tau = +\infty $ to $z = iy$. The inverse map is given by
\beq
 z_{\pm}(u) = -\frac{i}{u-1} \frac{1}{\ops} \left\{(1+u)\sst \pm\sqrt{(u-e^{i\sz})(u-e^{-i\sz})} \right\} ,
\eeq
where $u=e^\rho$. 

\FIGURE{\epsfig{file=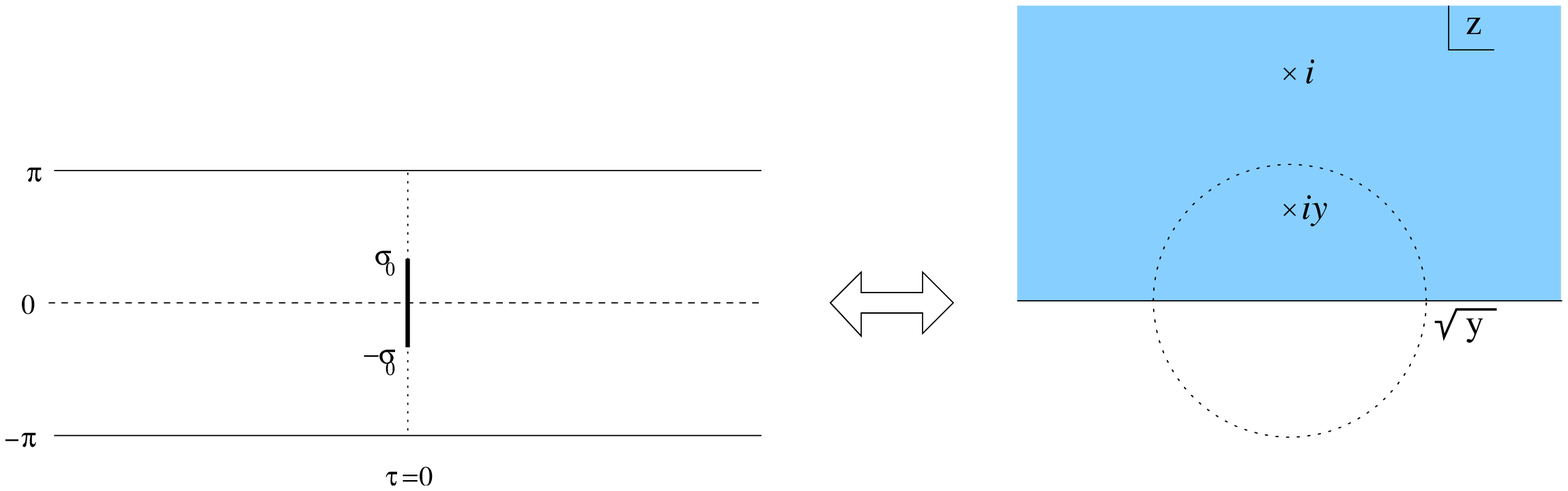, width=14cm}
\caption{Mandelstam map for the strip with a slit into the upper half plane. The point at $\tau=\pm\infty$ map to $z=iy$ and $z=i$ whereas
the slit maps to the real axis. The dotted line at $\tau=0$ and $\sz\le|\sigma|\le\pi$ maps to the half circle of radius $\sqrt{y}$.}
\label{fig:map}
}

 There is a sign ambiguity that has to be fixed. One can easily see that for $u\rightarrow 0$ we should take
$z_+$ and for $u\rightarrow\infty$ we should use $z_-$. Off course by appropriately defining the cut in the 
square root we obtain a function analytic in the region parameterized by $\rho$. In the regions
$u\rightarrow 0$ and $u\rightarrow\infty$ we can get a series expansion as:
\beqa
 z_1(u) &=& z_+(u) = \sum_{m=0}^{\infty} a_m u^m  ,\\
 z_2(u) &=& z_-(u) = \sum_{m=0}^{\infty} \frac{b_m}{u^m} . 
\label{zexp}
\eeqa
We already know that $a_0=i$ and $b_0=iy$. The other coefficients are given by
\beqa
 a_ m &=& \frac{2i}{m(\ops)} \sum_{l=1}^m \frac{m!}{(m+l)!} l (-)^l \left[\frac{\ops}{\cst}\right]^l \Plm \label{adef}\\
      &=& 2i\frac{\sst}{\ops} + \frac{i}{\ops}\sum_{l=1}^m \frac{1}{l} 
                  \left[\mathrm{P}_{l-2}(\cos\sz) -\cos\sz \mathrm{P}_{l-1}(\cos\sz) \right] ,\\ 
 b_ m &=& \frac{2i}{m(\ops)} \sum_{l=1}^m \frac{m!}{(m+l)!} l \left[\frac{\cst}{\ops}\right]^l \Plm \label{bdef}\\
      &=& -2i\frac{\sst}{\ops} + \frac{i}{\ops}\sum_{l=1}^m \frac{1}{l} 
                  \left[\mathrm{P}_{l-2}(\cos\sz) -\cos\sz \mathrm{P}_{l-1}(\cos\sz) \right] ,
\eeqa
 where we gave two alternative expressions. The first uses associated Legendre functions $\Plm$ and the other
Legendre polynomials. 

 Once we have mapped the problem to the upper half plane, computing the Green function is trivial. We obtain
\beq
 G(z,z') = \ln|z-z'| + \ei \ln|z-\bar{z}'| ,
\eeq
 where $\ei=1$ for Neumann boundary conditions and $\ei=-1$ for Dirichlet. We can define now four functions
\beqa
 \mathrm{N}^{rs}(u,u') = G(z_r(u),z_s(u')) - \delta^{rs} \ln|u-u'| ,
\eeqa
 with $r,s=1,2$ and we note that $\overline{z_r(u)} = -z_r(\bar{u})$. We also extracted the logarithmic singularity
as is conventional. Each function can be expanded as a power series in $u$ and $u'$ using (\ref{zexp}). After 
that we replace $u=e^{i\s}$, $u'=e^{i\s'}$. Thus, we obtain 
\beq
\NC^{rs}(u,u') = \sum_{m,n=\infty}^{\infty} \NNa e^{in\s} e^{im\s'} .
\eeq
 The coefficients $\NNa$ are precisely the Neumann coefficients we want to find. If we expand $G$ 
we obtain coefficients 
\beq
G^{rs}_{mn}= \NN - \frac{1}{2|n|} \delta_{n+m} \delta^{rs} .
\label{GN}
\eeq
The difference is just the expansion of $\ln|u-u'|$.

A straight-forward Taylor expansion can be used to compute a few of the coefficients. A more practical method of computation uses 
a trick. It is based on the fact that, by using the chain rule, we can obtain that
\beqa
 (\partial_\s + \partial_{\s'}) G &=& (1+\ei) \frac{z+\bar{z}+z'+\bar{z}'}{4(1-y)} \\ 
&& + \frac{\sqrt{y}(1+y)^2}{8(1-y)} \left[\left(\frac{1}{z+\sqrt{y}}+\frac{\ei}{\bar{z}+\sqrt{y}}\right)\left(\frac{1}{z'+\sqrt{y}}+\frac{\ei}{\bar{z}'+\sqrt{y}}\right)\right. \\
&& \left.-\left(\frac{1}{z-\sqrt{y}}+\frac{\ei}{\bar{z}-\sqrt{y}}\right)\left(\frac{1}{z'-\sqrt{y}}+\frac{\ei}{\bar{z}'-\sqrt{y}}\right)\right] ,
\label{Gder}
\eeqa
 the right hand side is a sum of terms which are factorized, namely are a product of a function of $z$ times a function of $z'$. We can then 
expand each of them and multiply the coefficients. The result is
\beq
 \NNa = -\frac{i}{8}\frac{(1+\ei)}{m+n} \left(a^{r}_m \delta_{n0} + a^s_n \delta_{m0} \right) 
  + \frac{1}{(m+n)\sin\sz} \mathrm{Im}\left(f^r_m f^s_n\right) ,
\label{Nfact}
\eeq
where
\beq
 a^{1,2}_m = 2i \pm \frac{i}{\sst} \sum_{l=1}^m \frac{1}{l} \left(\mathrm{P}_{l-2}(\cos\sz) -\cos\sz \mathrm{P}_{l-1}(\cos\sz)\right) , \ \ \ m>0 ,
\eeq
and $a^r_m = -a^r_{-m}$. The coefficients $f^r_m$ are given by
\beq
 f^1_{m>0} = -\bar{f}_m,  \ \ \ f^1_{m<0} = -\ei f_{-m}, \ \ \ f^2_{m\neq0} = -\ei f^1_m ,
\eeq
with
\beq
 f_{m>0} = -\frac{i}{m} e^{i\frac{\sz}{2}} \sum_{l=1}^m  \frac{(-i)^l m!}{(m+l)!} l \Plm .
\eeq
 Also,
\beqa
 f^1_0 &=& \half\left[ (1+\ei) (1-\sst) - i (1-\ei)\cst\right] ,\\
 f^2_0 &=& \half\left[ (1+\ei) (1+\sst) - i (1-\ei)\cst\right] .
\eeqa
 The result is valid when $m+n\neq 0$. If not we obtain
\beqa
 N^{11}_{m,-m} &=& -\frac{\ei}{2m^2}\left\{ \sum_{l=1}^{m-1}l\frac{m!(m-l-1)!}{(m+l)!(m-1)!} \Plm \mathrm{P}^l_{m-1}(\cos\sz)\right. \\
 && \left. + \frac{(-)^m m z_4^m}{(1-z_4)^{2m-1}} \left(\begin{array}{c} 2m-1\\m-1\end{array}\right) F(1,1-m;m+1;z_4)\right\} ,
\eeqa
and 
\beq
 N^{12}_{m,-m}+\ei N^{11}_{m,-m} = -\frac{1}{2|m|} ,
\eeq
 where $z_4=-\frac{\sin^2\frac{\sz}{2}}{\cos^2\frac{\sz}{2}}$ and $F$ denotes a hypergeometric function which in this case reduces to a polynomial.
We also have
\beq
N^{rs}_{i,00} = (1+\epsilon_i) \delta^{rs} \ln\left(\cos\frac{\sz}{2}\right) + \frac{1-\epsilon_i}{2}\ln\left(\sin\frac{\sz}{2}\right) .
\eeq

 To obtain the results we used the fact that $f_m$ and $a_m$ can be represented as contour integrals through  
\beqa
f_m(y) &=& \frac{i\,\sqrt{y}}{2\pi m}\oint_i dz \frac{1}{(z-\sqrt{y})^2} \left(\frac{(z+i)(z-iy)}{(z-i)(z+iy)}\right)^m ,\\
a_m(y) &=& -\frac{i}{2\pi m} \oint_i dz \left(\frac{(z+i)(z-iy)}{(z-i)(z+iy)}\right)^m . 
\eeqa
where the parameter $y$ was defined in eq.(\ref{ydef}). Here $a_m$ is the one defined in (\ref{adef}) and is related to $a^r_m$
due to eq.(\ref{Gder}) by
\beq
a^1_m =\frac{2}{1-y}\, a_m =\frac{1+\sst}{\sst}\, a_m,\ \  (m\ge 1)
\eeq  
 Using the method of residues we obtained the result in terms of sums that can be computed using
\beq
\sum_{p=0}^{m-l}\left(\begin{array}{c}m\\ p\end{array}\right)\left(\begin{array}{c}m\\ p+l\end{array}\right)
                \left(-\frac{1-\cos\sz}{1+\cos\sz}\right)^{p} =  \frac{2^m m!}{(m+l)!} \frac{(-1)^l}{(1+\cos\sz)^m} 
                \left(\frac{\sin\sz}{1+\cos\sz}\right)^l \mathrm{P}^l_m(\cos\sz)
\eeq
 To avoid confusion we recall the definition of associated Legendre functions:
\beq
\mathrm{P}^l_m(x) = (-1)^l (1-x^2)^{\frac{l}{2}} \frac{d^l}{dx^l} \mathrm{P}_m(x)
\eeq
where $P_m(x)$ are the standard Legendre polynomials. 

\subsection{Properties of the Neumann coefficients}

From the calculation we did one can find the following identities
\beqa
 N^{12}_{mn} + \ei N^{22}_{mn} &=& -\frac{1}{2|m|} \delta_{n,-m} + \frac{(1+\ei)}{2|m|} \delta_{n0} , \ \ \ m\neq 0 ,\\
 N^{12}_{mn} + \ei N^{11}_{mn} &=& -\frac{1}{2|n|} \delta_{m,-n} + \frac{(1+\ei)}{2|n|} \delta_{m0} , \ \ \ n\neq 0  .
\label{id1}
\eeqa
Other identities follow from the value of $G$ at $\tau=0$. For example, from the Dirichlet boundary condition,  we know that,
if $\ei=-1$, then $G(\s,\s') =0$ for $-\sz<\s<\sz$. In terms of Fourier components this can be written as
\beq
 \sum_n e^{in\s} G^{rs}_{nm} =0 , \ \ \ -\sz<\s<\sz, \ \ \ei=-1, 
\label{p3}
\eeq
 In the same way we can derive other identities. They follow from the value of $z$ when $\tau=0$. A simple calculation reveals that
$z_+(e^{i\phi})$ is real for $-\sz<\phi<\sz$ and is equal to $z_+(e^{i\phi})=\sqrt{y}e^{i\Phi}$ for $\sz<\phi< 2\pi-\sz$ with
$\cos\Phi=-\tan\frac{\sz}{2}\mathrm{cotan}\frac{\phi}{2}$. This means that the circle $\tau=0$ maps to the real axis and a circle of radius $\sqrt{y}$.
 The two sides of the slit map to different regions of the real axis corresponding to $|\mathrm{Re}(z)|$ greater or smaller than $\sqrt{y}$.
 If $z$ is real and $\ei=-1$ then $G(z,z')=0$ which is what we used to derive eq.(\ref{p3}). If $\ei=1$ then it seems that $\mathrm{Im}( \partial_z G)=0$ 
if $z$ is real, but there is a possible singularity if $z'$ is also real. Taking $z'=x'$ real and taking the limit $z\rightarrow x$ we get
\beq
 \mathrm{Im} (\partial_z G) = -\pi \delta(x-x') .
\label{a1}
\eeq
 Going to variables $\phi$ ,$\phi'$ and taking the Fourier transform we obtain
\beq
 \sum_n |n| e^{in\phi} \NN = -\half e^{-im\phi} \delta_{r\neq s} ,
\eeq
 where $\delta_{r\neq s} = 1-\delta_{rs}$. To get this result we used (\ref{GN}) and the fact that $\delta(x-x')$ on eq.(\ref{a1}) contributes only if $x$ can 
be equal to $x'$ which happens if $r=s$, namely both points are on the same side of the slit. 

 The next property is related to the values of the Green function for $\sz<\phi< 2\pi-\sz$. A lengthy but not difficult calculation reveals that
\beq
 \mathrm{Re}( \partial_\rho G )= -\frac{\pi}{2} \delta(\phi-\phi'), \ \ \ \ei=-1, \ \ u=e^{i\phi}, \ \ u'=e^{i\phi'}, \ \ \sz<\phi<2\pi-\sz .
\eeq
 Again, doing a Fourier transform and using (\ref{GN}) we obtain
\beq
 \sum_n |n| e^{in\phi} \NN = -\half e^{-im\phi} \delta^{r\neq s}, \ \ \ \ \ei=-1 .
\eeq
 The equivalent identity for $\ei=1$ is a little more difficult to derive. We start from the expression for $G$ and replace $z_+=\sqrt{y}e^{i\Phi}$.
We do the same for $z'$ but we need to consider two cases, when $-\sz<\phi'<\sz$ and when $\sz<\phi'<2\pi-\sz$. Then we can compute $G$ and obtain that
\beqa
 G^{11} &=& \ln\frac{\sin \frac{\sz}{2}}{\ops}+\ln|e^{i\phi}-e^{i\phi'}| - \ln |\frac{1-e^{-i\phi'}}{2}|
          - \ln |\frac{1-e^{-i\phi}}{2}| + \half (\ln z' +\ln \bar{z}') , \nonumber\\ && \nonumber\\
  \mbox{for} &\ & \ei=1, \ \ \sz<\phi<2\pi-\sz .
\label{a2}
\eeqa
We only computed $G^{11}$ since the others follow from the identities (\ref{id1}). From the relation 
between $\rho$ and $z$ one finds that
\beq
\frac{(u-1)^2}{u} = -\frac{4z^2(1-y)^2}{(1+z^2)(z^2+y^2)} .
\eeq
 Taking logarithms on both sides and taking the real part we find that
\beq
 2\ln z + 2\ln \bar{z} = G(z,i) + G(z,iy) -4\ln 2-4\ln(1-y) +4 \ln|1-u| -2\ln|u|  ,
\eeq
 where the Green function is taken with $\ei=1$. Doing a Fourier transform we find that
\beq
 \ln z + \ln \bar{z} = \sum_n \alpha_n e^{in\phi}, \ \ \ \alpha_{n\neq 0} = \NC^{11}_{n0} + \NC^{12}_{n0} - \frac{1}{|n|}, \ \  \ei=+1 .
\eeq
 Using this we can expand (\ref{a2}) to obtain
\beq
\sum_n e^{in\phi} \NC^{11}_{nm} = \half(\NC^{11}_{n0} + \NC^{12}_{n0}) , \ \ \ \sz<\phi<2\pi-\sz ,
\eeq
 \ie\ the function of $\phi$ on the left is actually a constant in that interval.

 We can summarize the properties of $\NN$ as
\begin{itemize}
\item{$\ei=-1$}\ (Dirichlet)
\beq
 \begin{array}{lcl}
  \sum_n e^{in\phi} \NN = \frac{1}{2|m|} e^{-im\phi} \delta^{rs}, & \ \ \ \ \ \ & -\sz<\phi<\sz \\ && \\
  \sum_n |n| e^{in\phi} \NN = -\frac{1}{2|m|} e^{-im\phi} \delta^{r\neq s}, & \ \ \ \ \ \ & \sz<\phi<2\pi-\sz ,
 \end{array}
\eeq
\item{$\ei=+1$}\ (Neumann)
\beq
 \begin{array}{lcl}
  \sum_{n=-\infty}^{\infty} |n| e^{in\phi} \NN = -\frac{1}{2} e^{-im\phi} \delta^{rs}, & \ \ \ \ \ \ & -\sz<\phi<\sz \\ && \\
  \sum_n     e^{in\phi} \NN = \frac{1}{2|m|}(2- e^{-im\phi}) \delta^{r\neq s} + (\NC^{11}_{n0} + \NC^{12}_{n0} )(\delta^{rs}-\half), 
                              && \sz<\phi<2\pi-\sz ,
 \end{array}
\eeq
\end{itemize}

 Other useful relations are:
\beqa
 N^{11}_{nm}(\ei=-1) &=& \sign(n)\, \sign(m)\,  N^{11}_{nm}(\ei=+1), \ \ \ \ (m,n\neq 0) \\
 N^{12}_{nm}(\ei=-1) &=& -\sign(n)\, \sign(m)\,  N^{12}_{nm}(\ei=+1), \ \ \ \ (m,n\neq 0) 
\eeqa
which can be used to derive
\beqa
\sum_{n\neq 0} \sign(n)  N^{11}_{nm}(\ei=-1) e^{in\s} &=& \half \sign(m) \left\{N^{12}_{m0}(\ei=+1)-N^{11}_{m0}(\ei=+1)\right\} \\
\sum_{n\neq 0} \sign(n)  N^{12}_{nm}(\ei=-1) e^{in\s} &=& \frac{1}{2m} e^{-im\s} + \half \sign(m) \left\{N^{12}_{m0}(\ei=+1)-N^{11}_{m0}(\ei=+1)\right\} \nonumber
\eeqa

 \subsection{Limit of large subindex}

It is useful to understand also various limits of the Neumann coefficients. One is their behavior for large values of one subindex. In view
of the expression (\ref{Nfact}) we only need to know the behavior of $a_m$ and $f_m$ for large $m$. These coefficients can be computed
by an integral in the complex plane which can be evaluated by a saddle point approximation. In terms of the parameter $y$ defined in eq.(\ref{ydef})
we have
\beqa
f_m(y) &=& \frac{i\,\sqrt{y}}{2\pi m}\oint_i dz \frac{1}{(z-\sqrt{y})^2} \left(\frac{(z+i)(z-iy)}{(z-i)(z+iy)}\right)^m ,\\
a_m(y) &=& -\frac{i}{2\pi m} \oint_i dz \left(\frac{(z+i)(z-iy)}{(z-i)(z+iy)}\right)^m . 
\eeqa
To understand the saddle point calculation it is useful to plot numerically the absolute value of the integrand as a function of $z$. This is a useful
aid to clarify the position of the saddle points. In any case, a standard calculation gives
\beqa
 f_m &\simeq_{m\rightarrow\infty}& \frac{\sqrt{\sin\sz}}{\sqrt{2\pi m}} e^{-i\frac{\pi}{4}} e^{im\sz} ,\\ && \nonumber\\
 a^1_m &\simeq_{m\rightarrow\infty}& 4i - \frac{2i}{\sqrt{2\pi}}\frac{1}{m^{\frac{3}{2}}} \frac{\cos\frac{\sz}{2}}{\sin\frac{\sz}{2}} 
         \frac{1}{\sqrt{\sin\sz}} \cos\left(m\sz-\frac{\pi}{4}\right) ,\\&& \nonumber\\
 a^2_m &\simeq_{m\rightarrow\infty}& \frac{2i}{\sqrt{2\pi}}\frac{1}{m^{\frac{3}{2}}} \frac{\cos\frac{\sz}{2}}{\sin\frac{\sz}{2}} 
         \frac{1}{\sqrt{\sin\sz}} \cos\left(m\sz-\frac{\pi}{4}\right) .
\eeqa
 Except for the constant part, the coefficients $a_m$ are subleading and can be ignored. For $m\rightarrow \infty$ the contribution comes
from $f_m$. 

\subsection{Limit $\sz\rightarrow 0$}

 Other limit of interest are $\sz\rightarrow 0$. Using standard properties of the Legendre polynomials one can derive that the functions
$f_{m>0}(\sz)$, $a^r_m(\sz)$ and $f^r_0(\sz)$ which enter in the formula for the Neumann coefficients behave as:
\beqa
 f_{m>0}(\sz) & \simeq & \frac{\sz}{2} \left(1+\frac{i\sz}{2}m-\frac{\sz^2}{4}\left(m^2+\frac{1}{6}\right)+\ldots \right), \\
 f^1_0(\sz) & \simeq & \half\left[(1+\ei)-i(1-\ei)\right]-\frac{1+\ei}{4}\sz + \frac{i}{16}(1-\ei)\sz^2 +\ldots ,\\
 f^2_0(\sz) & \simeq & \half\left[(1+\ei)-i(1-\ei)\right]+\frac{1+\ei}{4}\sz + \frac{i}{16}(1-\ei)\sz^2 +\ldots ,\\
 a^{1,2}_m(\sz) &\simeq & 2i\left(1\pm\frac{\sz}{2}m +\ldots\right) .
\eeqa 
 From here we can obtain the behavior of $\NN$. Assuming that $m,n>0$ we obtain that
\beq
 \NC^{rr}_{mn} \simeq -\frac{1}{8}\sz^2, \ \ 
 \NC^{rr}_{m,-n} \simeq -\frac{\ei}{8}\sz^2, \ \ 
 \NC^{rr}_{-m,-n} \simeq -\frac{1}{8}\sz^2, \ \ (\sz\rightarrow 0)
\eeq 
 which together with the properties (\ref{id1}) of $\NN$ determine completely the small $\sz$ behavior when $m,n \neq 0$. Note that, in particular, (\ref{id1})
implies that
\beq
 N^{12}_{m,-m} \simeq -\frac{1}{2|m|} + \frac{\sz^2}{8} + \ldots ,
\eeq
 has an order zero contribution. In the case that one subindex is zero we get
\beqa
N^{11}_{m0} & \simeq & \frac{1}{2|m|} + \frac{1+\ei}{4} \sz - \frac{\sz^2}{16} \left[(1+\ei)+|m|(1-\ei)\right] + \ldots  ,\\ 
N^{12}_{m0} & \simeq & \frac{1}{2|m|} + \frac{1+\ei}{4} \sz - \frac{\sz^2}{16} \left[-(1+\ei)+|m|(1-\ei)\right] + \ldots ,\\ 
N^{21}_{m0} & \simeq & \frac{1}{2|m|} - \frac{1+\ei}{4} \sz - \frac{\sz^2}{16} \left[-(1+\ei)+|m|(1-\ei)\right] + \ldots ,\\ 
N^{22}_{m0} & \simeq & \frac{1}{2|m|} - \frac{1+\ei}{4} \sz - \frac{\sz^2}{16} \left[(1+\ei)+|m|(1-\ei)\right] + \ldots .
\label{Nsz0}
\eeqa
 Where in all cases we have expanded to the order that is necessary for the main part of the paper.

\commentout{ 
 We can also write this as
\begin{itemize}
\item{$\ei=-1$}
\beq
 h_{mn} \NC^{rs}_{nk}= \frac{1}{2|k|} h_{m,-k} \delta^{rs} 
\eeq
\beq
 (1-h)_{mn} |n| \NC^{rs}_{nm} = -\half (1-h)_{m,-k}\delta^{r\neq s}
\eeq
\item{$\ei=+1$}
\beq
 h_{mn} |n| \NC^{rs}_{nk} = -\half h_{m,-k} \delta^{rs} 
\eeq
\beq
 (1-h)_{mn} \NC ^{rs}_{nk} = -\frac{1}{2|k|} h_{m,-k} \delta^{r\neq s} + \left[\frac{1}{|k|}\delta^{r\neq s}+ 
  (\NC^{11}_{n0} + \NC^{12}_{n0} )(\delta^{rs}-\half)\right]h_m
\eeq
\end{itemize}
}


\end{document}